\documentclass[a4paper,fleqn,usenatbib]{mnras}

\usepackage{newtxtext,newtxmath}

\usepackage[T1]{fontenc}
\usepackage{ae,aecompl}

\usepackage{graphicx}	
\usepackage{amsmath}	
\usepackage{amssymb}	

\usepackage{upgreek}

\usepackage{latexsym}
\usepackage{wasysym}

\usepackage{float} 
\usepackage{wrapfig} 
\usepackage{color}
\usepackage{caption}
\usepackage{subcaption}
\usepackage{pdflscape}

\usepackage{soul}

\newcommand{\irs}{Oph~IRS~48}
\newcommand{\hd}{HD~169142}
\newcommand{\chisquare}{$\chi^2_{\rm shot}$}

\newcommand{\radmc}{\texttt{RADMC3D}}
\newcommand{\emcee}{\texttt{emcee}}
\newcommand{\radmcpy}{\texttt{RADMC3DPy}}
\newcommand{\msolar}{\mathrm{M}_\odot}
\newcommand{\rsol}{\mathrm{R}_\odot}
\newcommand{\au}{\mathrm{AU}}
\newcommand{\pc}{\mathrm{pc}}
\newcommand{\K}{\mathrm{K}}
\newcommand{\degree}{^\circ}
\newcommand{\marcsec}{mas}

\title[Tiny Grains in Gaps]{Tiny Grains Shining Bright in the Gaps of Herbig Ae Transitional Discs}

\author[Eloise K. Birchall et al.]{
Eloise K. Birchall$^{1}$,\thanks{E-mail: eloise.birchall@anu.edu.au(EKB)}
Michael J. Ireland$^{1}$,
Christoph Federrath$^{1}$,
John D. Monnier$^{2}$,
\newauthor
Stefan Kraus$^{3}$,
Matthew Willson$^{3}$,
Adam L. Kraus$^{4}$, 
Aaron Rizzuto$^{4}$,
\newauthor
Matthew T. Agnew$^{5}$,
Sarah T. Maddison$^{5}$
\\
$^{1}$Research School of Astronomy and Astrophysics, Australian National University, Canberra, 2611,  ACT, Australia\\
$^{2}$Department of Astronomy, University of Michigan, Ann Arbor, Michigan 48109, USA\\
$^{3}$University of Exeter, Astrophysics Group, School of Physics, Stocker Road, Exeter, EX4 4QL, UK\\
$^{4}$Department of Astronomy, The University of Texas at Austin, Austin, TX 78712, USA\\
$^{5}$Centre for Astrophysics and Supercomputing, Swinburne University of Technology, Hawthorn, Victoria 3122, Australia\\
}

\date{Accepted XXX. Received YYY; in original form ZZZ}

\pubyear{2019}

\begin{document}
\label{firstpage}
\pagerange{\pageref{firstpage}--\pageref{lastpage}}
\maketitle

\begin{abstract}
This work presents a study of two Herbig Ae transitional discs, \irs~and \hd; which both have reported rings in their dust density distributions.
We use Keck-II/NIRC2 adaptive optics imaging observations in the L' filter ($3.8$\,\micron) to probe the regions of these discs inwards of $\sim20\,\au$ from the star.
We introduce our method for investigating these transitional discs, which takes a forward modelling approach: making a model of the disc (using the Monte Carlo radiative transfer code \radmc), convolving it with point-spread functions of calibrator stars, and comparing the convolved models with the observational data. 
The disc surface density parameters are explored with a Monte Carlo Markov Chain technique.
Our analysis recovers emission from both of the discs interior to the well known optically thick walls, modelled as a ring of emission at $\sim15\,\au$ in \irs, and $\sim7\,\au$ for \hd, and identifies asymmetries in both discs. 
Given the brightness of the near-symmetric rings compared to the reported companion candidates, we suggest that the reported companion candidates can be interpreted as slightly asymmetric disc emission or illumination.

\end{abstract}




\section{Introduction}

Transitional discs are a subset of protoplanetary discs that have a region which is depleted in dust (but not necessarily depleted in gas), known as a gap or hole. 
These objects are a subset of \citet{lada_star_1987} Class II objects, with flat or declining mid-infrared (mid-IR) excesses in their spectral energy distributions (SEDs).
Giant planets that have formed in protoplanetary discs are predicted to carve out gaps or holes, producing structures that might be observed in transitional discs \citep{strom_circumstellar_1989, skrutskie_sensitive_1990, marsh_evidence_1992, marsh_spectral_1993}. 
It has also been suggested that the gaps in protoplanetary discs could be caused by multi-planet systems \citep{dodson-robinson_transitional_2011}, however the prevalence of systems with 3 or more Jupiter mass planets is very low \citep[$<1$\%,][]{han_exoplanet_2014}, so this is unlikely to be the only cause of transitional disc structure.
By direct imaging of transitional discs, we can look for signs of planet formation or other disc evolution such as dust asymmetries, dust depleted regions, gaps, rings, and holes.

The transitional discs studied in this work are \irs~and \hd. 
Both are Herbig Ae stars known to have polycyclic aromatic hydrocarbons (PAHs) in their discs.
Information on these objects is summarised in Table \ref{tab:stars}.
Both objects are known to have structure in their outer discs with rings of emission at $50-60\,\au$ identified from observations at mm wavelengths \citep[e.g.][]{geers_spatial_2007, brown_matryoshka_2012, van_der_marel_major_2013, quanz_gaps_2013, maaskant_polycyclic_2014}.

There are several objects for which there are detections of dust within the regions thought to be depleted of CO.
This is the case for V1247 Orionis, for which the presence of carbon-rich dust inside the gap region has been found through near-IR observations \citep{kraus_resolving_2013}.
Previous modelling of infrared spectra of both \irs~and \hd~have indicated the likely presence of PAHs within the gas gap regions \citep[regions depleted in CO;][]{geers_spatially_2007, maaskant_polycyclic_2014,seok_polycyclic_2016}.

\begin{table}
\caption{Parameters of \irs~and \hd}
\begin{center}
\footnotesize{{\begin{tabular}{ccc}\hline
\multicolumn{3}{c}{\irs} \\
\hline 
Right Ascension & 16h 27m 37.190s & 1  \\
Declination &$-24\,^\circ 30\,\arcmin 35.03\,\arcsec$ & 1 \\
Alternate Names & WLY 2-48, & 1\\
 &  2MASS J16273718-2430350, & 1\\
 &  YLW 46 & 1\\
Stellar Type & A0 & 2, 3 \\
Distance to object & $134.4\pm{2.2}\,\pc$ & 4 \\
W1 magnitude & 5.786 & 5 \\
\hline
\hline
\multicolumn{3}{c}{\hd} \\
\hline
Right Ascension & 18h 24m 29.779s & 1 \\
Declination & $-29\,^\circ 46\,\arcmin 49.37\,\arcsec$ & 1 \\
Alternate Names & MWC 925 & 1 \\
Stellar Type & A5 & 6 \\
Distance to object & $114.0\pm{0.8}\,\pc$ & 4 \\
W1 magnitude & 6.203 & 5 \\
 \hline

\end{tabular}}}
\begin{minipage}{\linewidth}
\vspace{0.1cm}
\footnotesize{\textbf{Notes.} 1. SIMBAD: simbad.u-strasbg.fr/; 2. \citet{mcclure_evolutionary_2010}; 3. \citet{brown_30_2012}; 4.\citet{gaia_collaboration_gaia_2018}; 5. WISE All-Sky Catalog; 6. \citet{seok_dust_2016}}.
\end{minipage}
\end{center}
\label{tab:stars}
\end{table}

\irs~is located in the $\rho$ Ophiuchus star forming region, as catalogued by \citet{elias_infrared_1978} and confirmed by \citet{wilking_iras_1989}.
The average distance to the cloud's core was $120.0^{+4.5}_{-4.2}\,\pc$ found by \citet{loinard_preliminary_2008} using Very Long Baseline Array (VLBA) data. 
The parallax for \irs~as given by Gaia DR2  \citep{gaia_collaboration_gaia_2018} is $7.44\pm0.12$\,millisecond of arc (mas), corresponding to a distance of $\sim134.4\pm2.2\,\pc$, and this is the distance adopted in this work.

The outer disc of \irs~is known to have a strong asymmetric feature in the millimetre-sized grains.
The asymmetry in the dust of the outer disc (at a deprojected distance of $67\,\au$ from the star) was discovered with ALMA sub-millimetre observations and indicates a separation of the micron- and millimetre-sized dust grains \citep{van_der_marel_major_2013}.
There is also a separation of the very small grains that do not follow the larger micron sized dust, as reported as unresolved emission in  \citet{geers_spatial_2007}.
Later observations found that centimetre-sized dust grains are further concentrated in the region of the millimetre-grains \citep{van_der_marel_concentration_2015}, consistent with being caused by a vortex induced by a pressure or density gradient generating a dust trap, which may have been generated by a companion \citep{van_der_marel_major_2013, van_der_marel_concentration_2015}.

Inside the $67\,\au$ asymmetric ring, the structure of the \irs~disc becomes more complicated.
It is thought that there is a depletion of dust within $67\,\au$, given that there is a no excess at wavelengths short of 10\,\micron~in the SED \citep{maaskant_identifying_2013}.
However, where these depletions are seen can depend on the wavelength of the observations, as different materials (gas, dust, PAHs, large or small grains) are traced by different wavelengths \citep[e.g.][]{brown_matryoshka_2012}.
The disc is considered to be mostly depleted of dust within $\sim23\,\au$ \citep{bruderer_gas_2014}, where there is a wall, but the flux deficit caused by depletion depends strongly on the wavelength of observation. 
\irs~is still accreting \citep[accretion rate $\sim10^{-9}\,\msolar\,\mathrm{yr^{-1}}$,][]{salyk_measuring_2013}, and SED modelling also confirms the presence of inner-disc material \citep[e.g.][]{bruderer_gas_2014}.

\hd~was previously thought from optical photometry to be located at a distance of $145\,\pc$ \citep{sylvester_optical_1996}, but the parallax was measured with Gaia \citep{gaia_collaboration_gaia_2018} to be $8.77\pm 0.06\,$\marcsec, corresponding to a distance of $114.0\pm 0.8\,\pc$. 
We adopt the new distance in this work, and have adjusted the linear radii of the previously noted features to correspond to the new distance.

\hd~was reported as having circumstellar material by \citet{walker_cool_1988}, and subsequent studies have found that this disc has structure, including an inner cavity and rings \citep[e.g.][]{panic_gas_2008, honda_mid-infrared_2012, quanz_gaps_2013, osorio_imaging_2014, seok_dust_2016, monnier_polarized_2017, fedele_ALMA_2017, macias_imaging_2017}.
The most commonly modelled structure for \hd~thus far includes a small inner rim \citep{panic_gas_2008, honda_mid-infrared_2012}, with an optically thick ring at $\sim20\,\au$  \citep{osorio_imaging_2014, honda_mid-infrared_2012, panic_gas_2008}, and a gap from $31\,\au$ to an optically thick wall at $55\,\au$ \citep{quanz_gaps_2013}.

It is possible that the gaps in the disc are caused by planet-disc interactions.
\citet{biller_enigmatic_2014} and \citet{reggiani_discovery_2014} suggested from L' (3.8\,\micron) observations using a coronagraph that a potential companion is located at a distance of either $\sim 13\,\au$  \citep[$110\pm{30}\,$mas, position angle of $0\pm{14}\,\degree$,][]{biller_enigmatic_2014} or $\sim 18\,\au$ \citep[156$\pm{32}\,$mas, position angle of $7.4\pm{11.3}\,\degree$,][]{reggiani_discovery_2014} from the star (interior to the $\sim20\,\au$ ring). 
The potential companion has an L' apparent Vega magnitude of $12.2\pm0.5$\,mag \citep{reggiani_discovery_2014}, which corresponds to a contrast of 6.4\,mag \citep{biller_enigmatic_2014}, or $6.5\pm0.5$\,mag \citep{reggiani_discovery_2014}.
These studies both used observations taken with the Very Large Telescope using NACO, and the coronograph available for this instrument.
Both studies used similar methods to detect the candidate companion, which included angular differential imaging (ADI) that self-subtracted any azimuthally symmetric structures.
Other studies investigate whether this possible companion, and potentially a second candidate companion further out in the disc are causing the structural features seen in the disc \citep{fedele_ALMA_2017, kanagawa_mass_2015}.

There have been few previous imaging studies of these discs at wavelengths shorter than $\sim$8\,$\mu$m -- wavelengths most sensitive to structures inside the known rings and where planet formation signatures, including those of circumplanetary accretion discs \citep{Zhu15}, could be seen. 
The key reason for this gap is the higher contrasts and angular resolutions needed for shorter wavelengths, meaning that specialised techniques are required, such as ADI \citep{reggiani_discovery_2014}, or aperture mask interferometry \citep{Schworer17}. 

It is possible that there is a companion in each of the discs, clearing the gaps and driving the asymmetries \citep[e.g.][]{van_der_marel_major_2013, biller_enigmatic_2014, reggiani_discovery_2014}.
To investigate whether there is a planet inside the gas gaps of these discs at $\sim20\,\au$ we present our observations and analysis of both discs as follows.
Section \ref{sec:obs} discusses the observations used for this work, which differs from previous studies as we are more sensitive to extended structures than aperture mask interferometry, but are still sensitive to circularly symmetric features, unlike ADI.
Section \ref{sec:redux} discusses evidence for significant emission inside the previously inferred walls.
Sections~\ref{sec:models} and \ref{sec:synth} outlines the computational methods used in this paper and tests them on a synthetic data set.
Section \ref{sec:results} describes the resulting best fit physical disc models.
Section \ref{sec:conclusion} summarises our results in the context of the field, and discusses possible future work in this area.

\section{Observations and Data Reduction} \label{sec:obs}

\subsection{Observations}

Observations were acquired with the NIRC2 instrument on Keck II over three observing runs (June 2014, 2015, 2016).
All observations used in this work were taken with the L' filter of the NIRC2 camera, using a `large hex pupil', a $512\times512$ subarray, and a two-point dither mode separated by $\sim$3.5'', utilising the top left and bottom right quadrants of the detector. 
These observations did not use aperture masking or coronography. 
Table \ref{tab:obs} summarises these observations, including total exposure times and point spread function (PSF) calibrators. 
We use both natural and laser guide star adaptive optics (AO) in this work (because \hd~is bright enough to not need the laser, and \irs~is fainter). Our 2014 and 2016 observations of \hd~used natural guide star AO and the 2015 and 2016 observations of \irs~used laser guide star AO.

The calibrators for \hd~were chosen specifically for that object, while \irs~was calibrated against other Class~II objects that formed a survey of the Ophiuchus star forming region for accreting exoplanets. 
These calibrators were expected to have bright, unresolved inner discs, with cross-calibration between different calibrators used to check for any measurably resolved objects. 
No potential calibrators were eliminated. 

\begin{table*}
\caption{Observations.} 
\label{tab:obs}
\begin{center}
{\footnotesize\makebox[\textwidth][c]{\begin{tabular}{ccccccccccc}\hline
Target & Date &  T int  & Coadds & Exposure  & Frames & Visits & Seeing & Airmass  & Calibrators\\
 & & (s) & & Time (s) & & & ('')&  & (Visits) \\
\hline 

\hd & 10 June 2014  & 0.053 & 200 & 10.6 & 8, 4 & 2 & 0.70 & 1.61  & HD 167666 (1), HD 170768 (2) \\
\irs & 23 June 2015  & 0.1 & 100 & 10 & 18 & 1 & 0.30 & 1.40  & Elia 2-24 (1), Elia 2-26 (1), \\
& & & & & & & & & WSB 52 (1) \\
\irs & 16 June 2016  & 0.2 & 100 & 20 & 8, 10 & 2 & No Data & 1.52, 1.42 &  GSS 37 (2), DoAr 32 (2), \\
 &  & &  & &  & &  & &  DoAr 24 (2), WSB 12 (2),  \\
 &  & &  & &  & &  & &   DoAr 33 (2), WSB 52 (1),  \\
 &  & &  & &  & &  & &   WSB 52/[WMR2005] 2-30 (1),  \\
 &  & &  & &  & &  & &   [MMG98] RX J1622.9-2326 (2) \\
\hd & 17 June 2016  & 0.2 & 100 & 20 & 10 & 1 & 0.95 & 1.56  & HD 167666 (1) \\

\hline

\end{tabular}}}

\begin{minipage}{\linewidth}
\vspace{0.1cm}
\footnotesize{\textbf{Notes.} All of these observations were taken in the L' filter using the NIRC2 instrument on the Keck II telescope. 

Column~1: Target name. 
Column~2: Date the observations were taken. 
Column~3: Integration time for each coadd in seconds. 
Column~4: Number of coadds (snapshots that make up the final image).
Column~5: Total exposure time in seconds, integration time multiplied by coadds.
Column~6: Frames taken each visit. 
Column~7: Number of visits to that target that night. 
Column~8: The seeing is taken as the mean seeing for that night from the Canada-France-Hawaii Telescope seeing monitor and is in the V filter. 
Column~9: Airmass the target was observed through. 
Column~10: Calibrators used (Number of visits to calibrator).} 
\end{minipage}

\end{center}
\end{table*}

\subsection{Data Reduction} \label{sec:datred}

Each image was corrected for detector nonlinearity using the algorithm from the IDL program {\tt linearize\_nirc2.pro} \footnote{http://www.astro.sunysb.edu/metchev/ao.html}, and then divided by a mean dome flat field.
Images taken with the star at the opposite dither position were used as an estimate of sky, which was subtracted from each image.
Bad pixels were primarily identified by searching for outliers in mean pixel value or variance in a sequence of dark frames or flat frames. Additional bad pixels or cosmic rays were found by Fourier transforming each data frame, masking out spatial frequencies below Nyquist, inverse transforming and searching for significant peaks. 
The bad pixels were corrected using the algorithm in \citet{ireland_phase_2013}, which involves finding the pixel value that minimises Fourier power above the Nyquist frequency. 
Images were also centred to the peak value (which is typically the centre of the star) and cropped to $128\times128$ pixels, to remove edge effects.

When making the set of cleaned images to be used for subsequent analysis, images extincted by cloud by more than a factor of 2 or with Strehl more than $\sim$20\% below the maximum were rejected. 
For the 2014 \hd~data, only one of the 12 target images was rejected.
For the 2015 \irs~data, three of the 18 target images were rejected.
For the 2016 \hd~data, two of the 10 target images were rejected.
For the 2016 \irs~data, six of the 18 target images were rejected.
The 2016 data were taken through cloud, and for all data sets we began data acquisition before the low-bandwidth wavefront sensor showed low errors. 
All raw data are publicly available in the Keck observatory archive, and we have made the cleaned data cubes\footnote{http://www.mso.anu.edu.au/$\sim$mireland/Birchall18/} available.


\begin{figure*}
\centering
\resizebox{\textwidth}{!} {\begin{tabular}{@{}p{0.34\textwidth}p{0.34\textwidth}@{}}
\includegraphics[width=.34\textwidth]{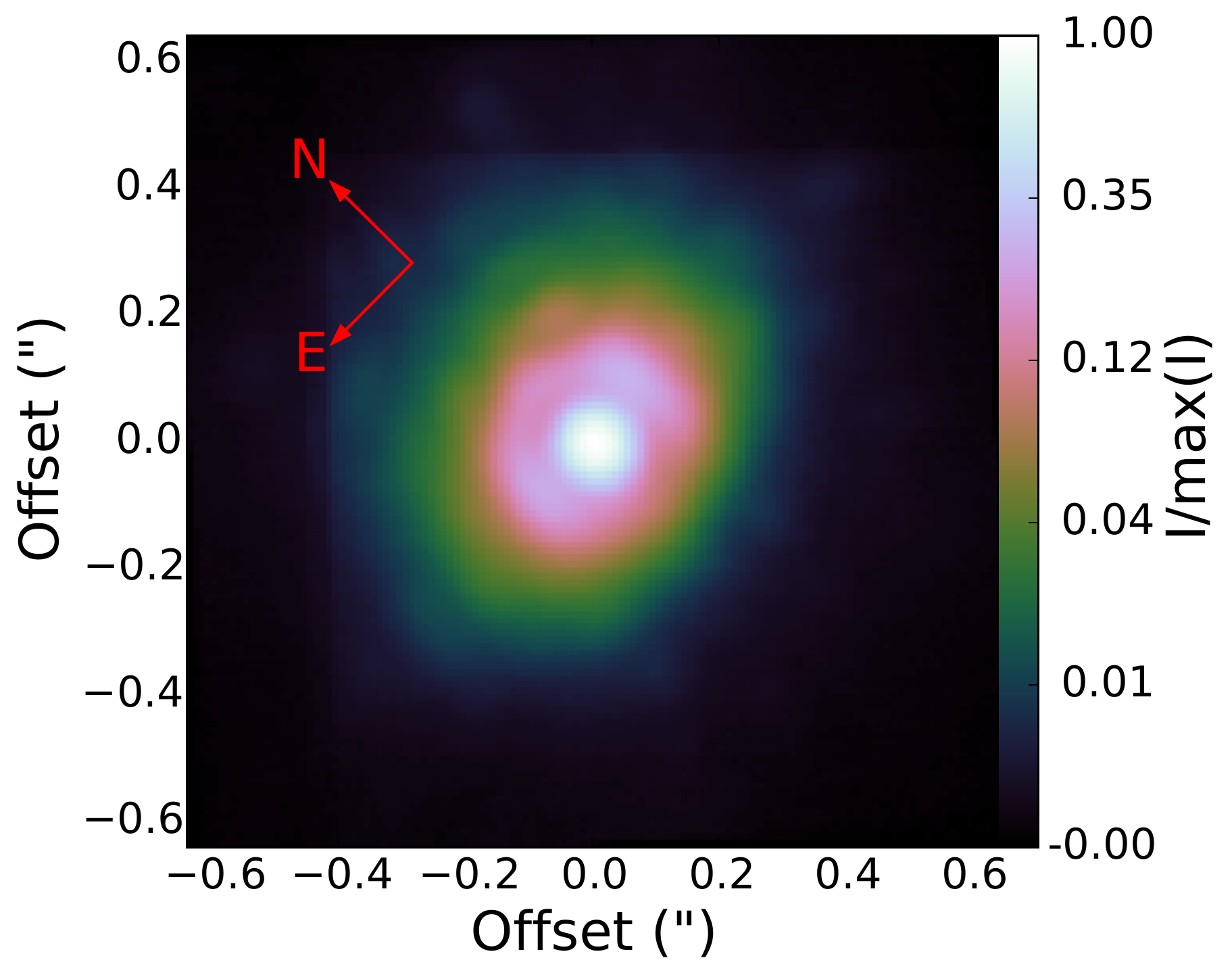} &
\includegraphics[width=.34\textwidth]{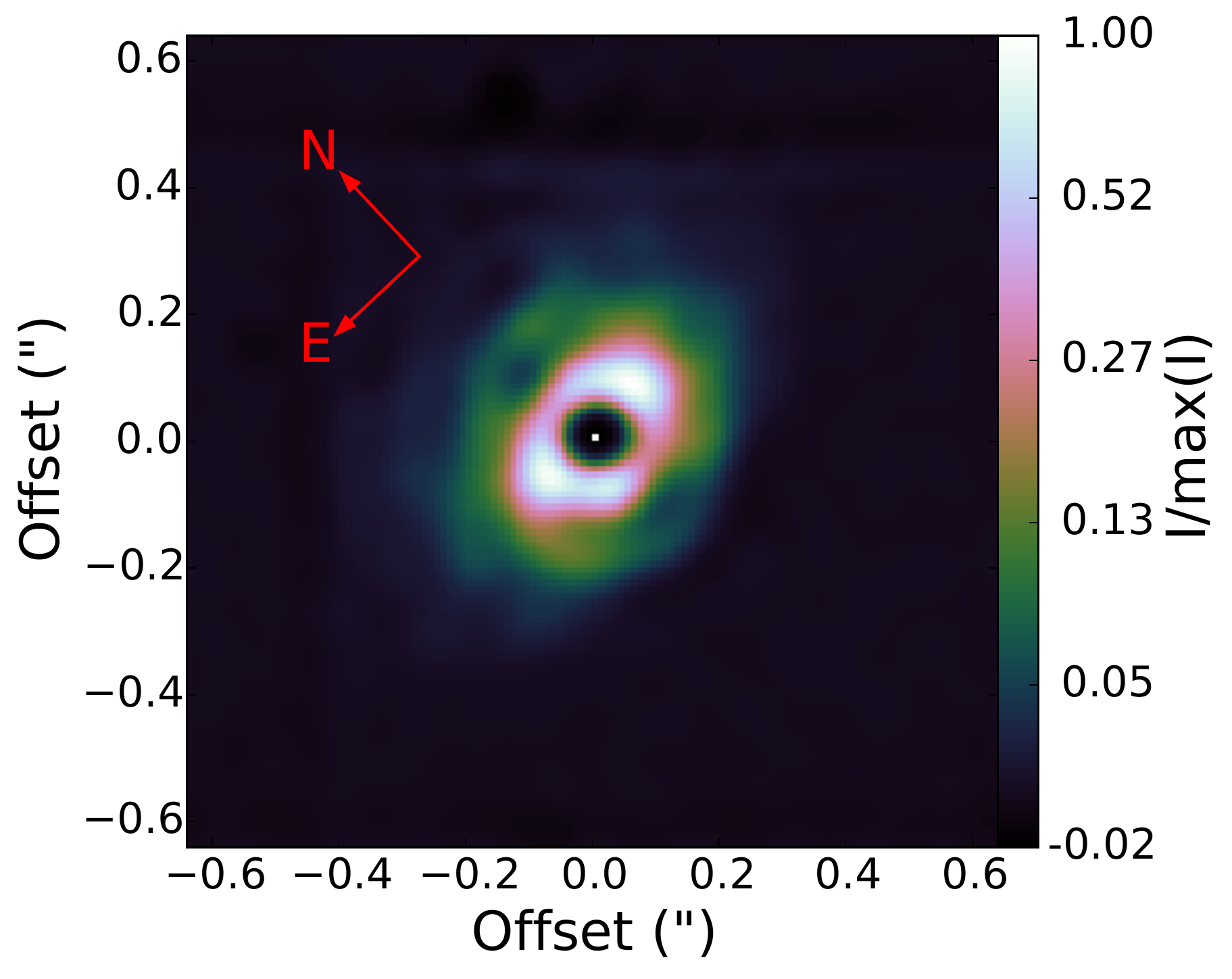} \\[-2.5pt]
\end{tabular}}
\caption{Left: Reduced data for \irs. Right: Deconvolved, point source subtracted image of \irs. The central bright point shows where the star would be. Arrows show the direction of north and east. Images have arcsinh colour scaling, and are normalised to the maximum intensity of the image.}
  \label{irs_data}
\end{figure*}

\begin{figure*}
\centering
\resizebox{\textwidth}{!} {\begin{tabular}{@{}p{0.34\textwidth}p{0.34\textwidth}@{}}
\includegraphics[width=.34\textwidth]{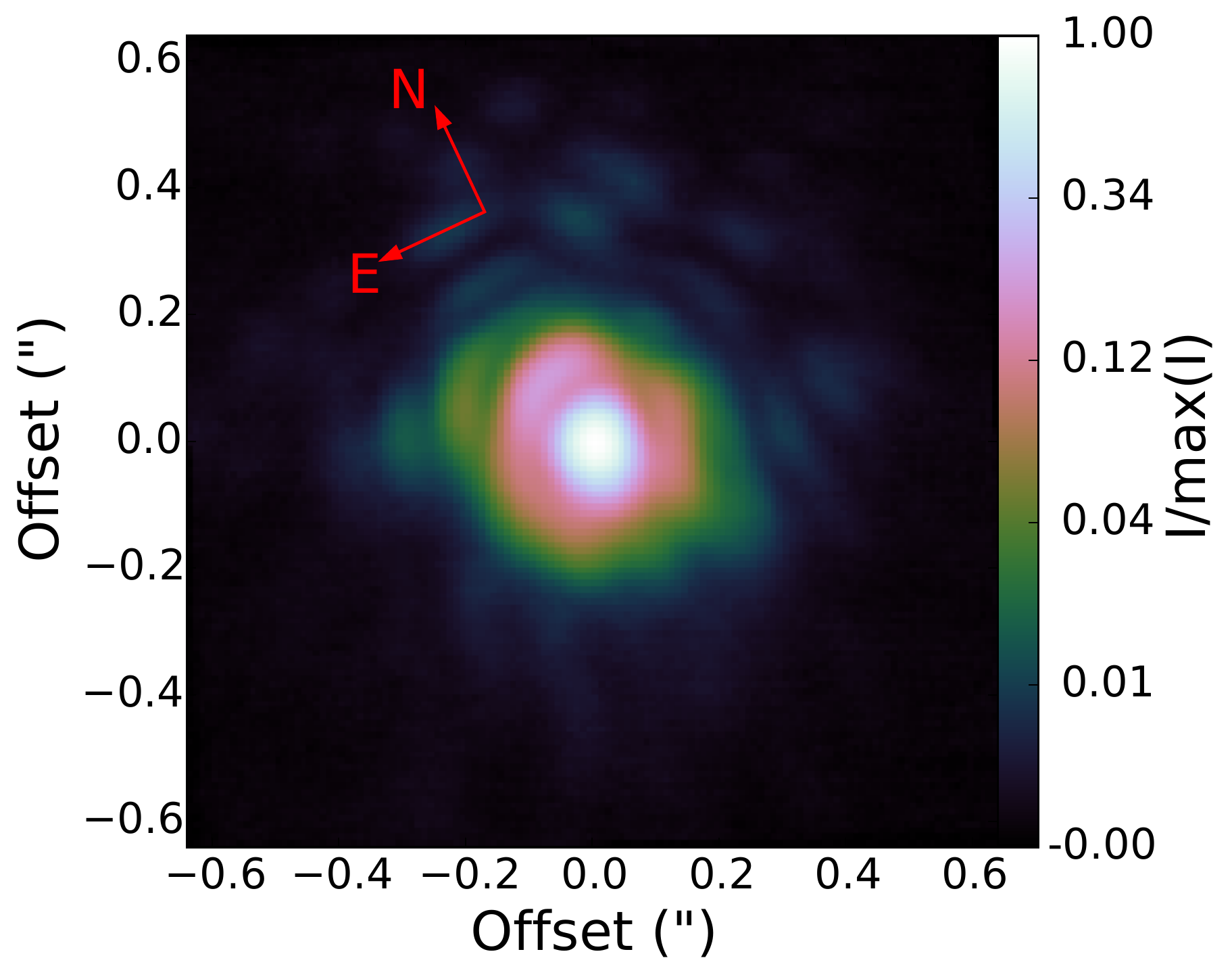} &
\includegraphics[width=.34\textwidth]{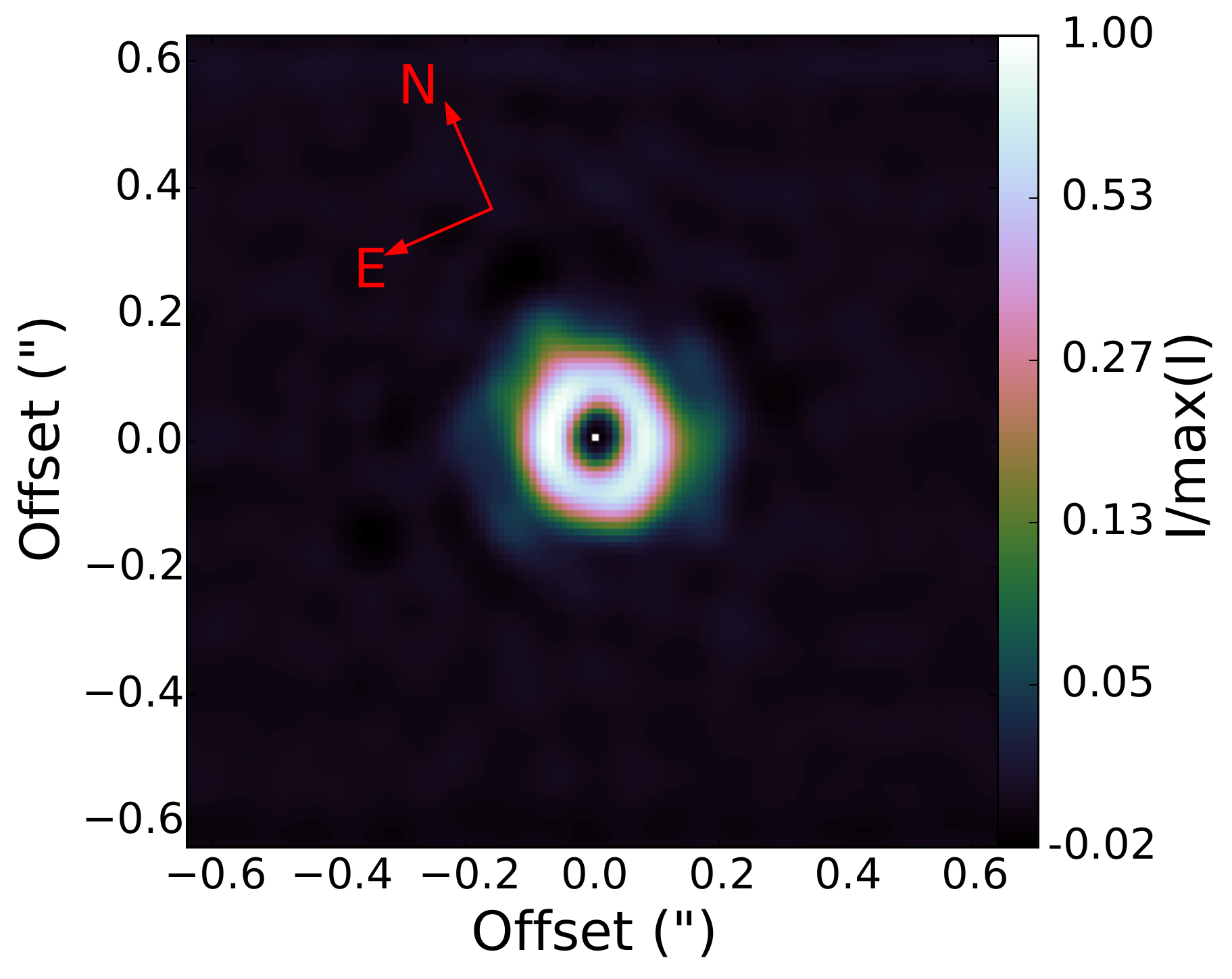} \\[-2.5pt]
\end{tabular}}
\caption{Left: Reduced data for \hd. Right: Deconvolved, point source subtracted image of \hd. The central bright point shows where the star would be. Arrows show the direction of north and east. Images have arcsinh colour scaling, and are normalised to the maximum intensity of the image. }
  \label{hd_data}
\end{figure*}

\section{Resolved Disc Emission} \label{sec:redux}

\subsection{Richardson-Lucy Deconvolution} \label{sec:rldeconv}

The Richardson-Lucy method of deconvolution is an iterative deconvolution method based on finding the mostly likely image that fits a data set given a known PSF. It was developed independently by \citet{richardson_bayesian-based_1972} and \citet{lucy_iterative_1974}.
We performed a Richardson-Lucy deconvolution on our reduced data, in order to isolate the resolved structure in the discs, and as a simple first-pass imaging algorithm to inspire physical model-fitting.

We used 50 iterations of the Richardson-Lucy algorithm, and our initial model was a point-source.
We used this algorithm on all target-image -- calibrator-image pairs.
However, we only considered the calibrator image that, when convolved, had the smallest root-mean-square (RMS) with respect to the target star image (i.e. the one that returned the image most similar to each target image).
We chose to use 50 iterations because at this point the RMS between the target and calibrator images is stable, and not decreasing dramatically with more iterations. 
Also, with more iterations, the deconvolution noticeably over fits the data.

One should note that the point-source initial model for our implementation of the Richardson-Lucy algorithm has an effect of regularising the final model to be a point source with resolved structure, rather than a marginally-resolved central source with structure.
A point source with a disc is a suitable physical approximation for our deconvolved images as the star and its inner ring of material ($<1\,\au$) is unresolved. 

The deconvolved images for \irs~and \hd~are shown in the right panels of Figures \ref{irs_data} and \ref{hd_data}. In both cases, a resolved ring of emission is clearly visible at radii within the 
previously published transitional disc holes ($\sim23\,\au$~for \irs, and $\sim20\,\au$~for \hd).

\subsection{Fourier Analysis}

\begin{figure}
   \centering
	\includegraphics[width=0.5\textwidth]{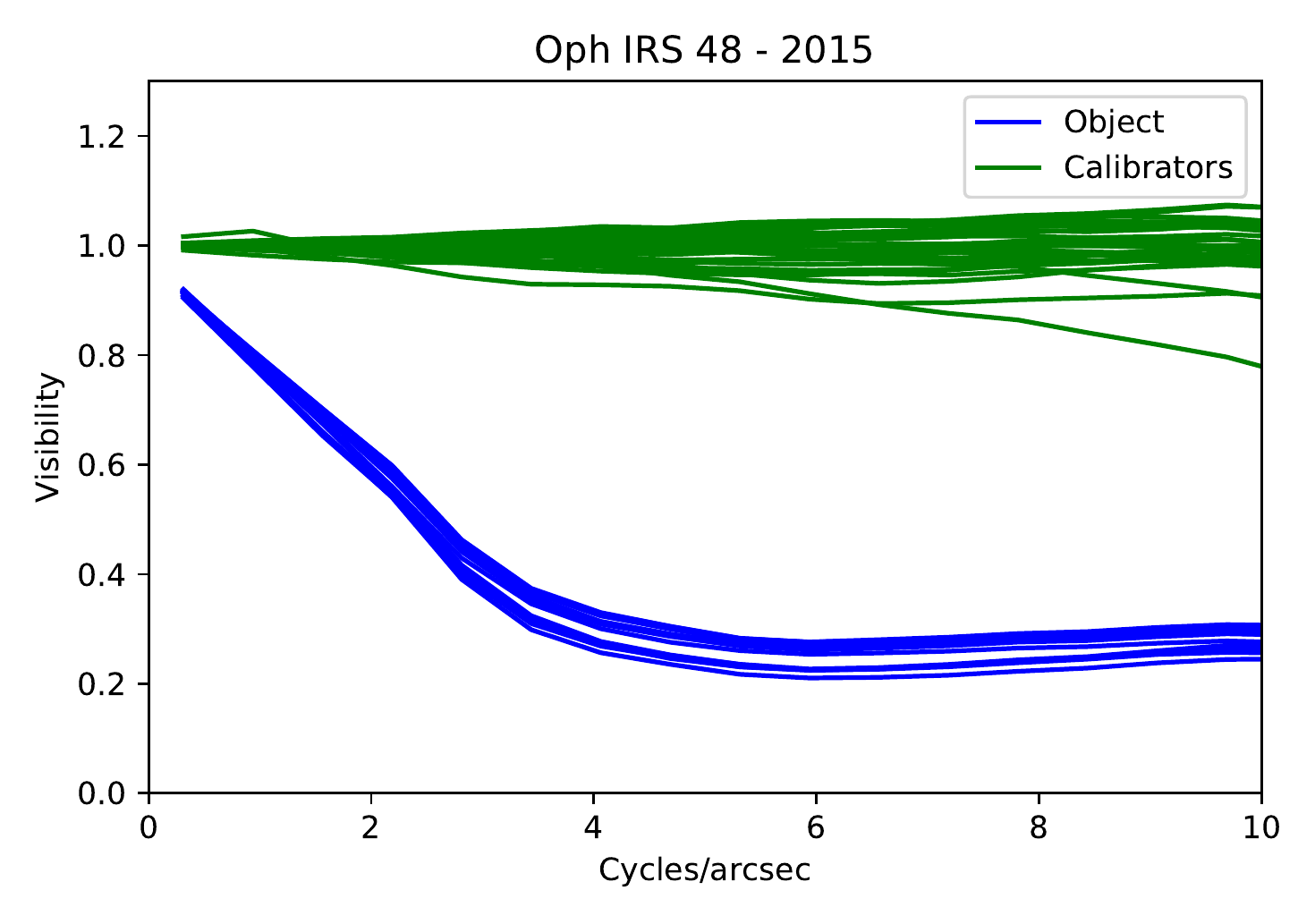}
                \caption{The azimuthally averaged power spectrum for Oph~IRS~48 and calibrators for the 2015 epoch. Each
                line represents one saved 10\,s exposure. In this domain, it is both very clear that the target is well-resolved and that 
                the calibrators have stable visibilities (square root of power) with a standard deivation of  $\lesssim$5\%.}
  \label{figFourierIRS48}
\end{figure}

\begin{figure}
   \centering
	\includegraphics[width=0.5\textwidth]{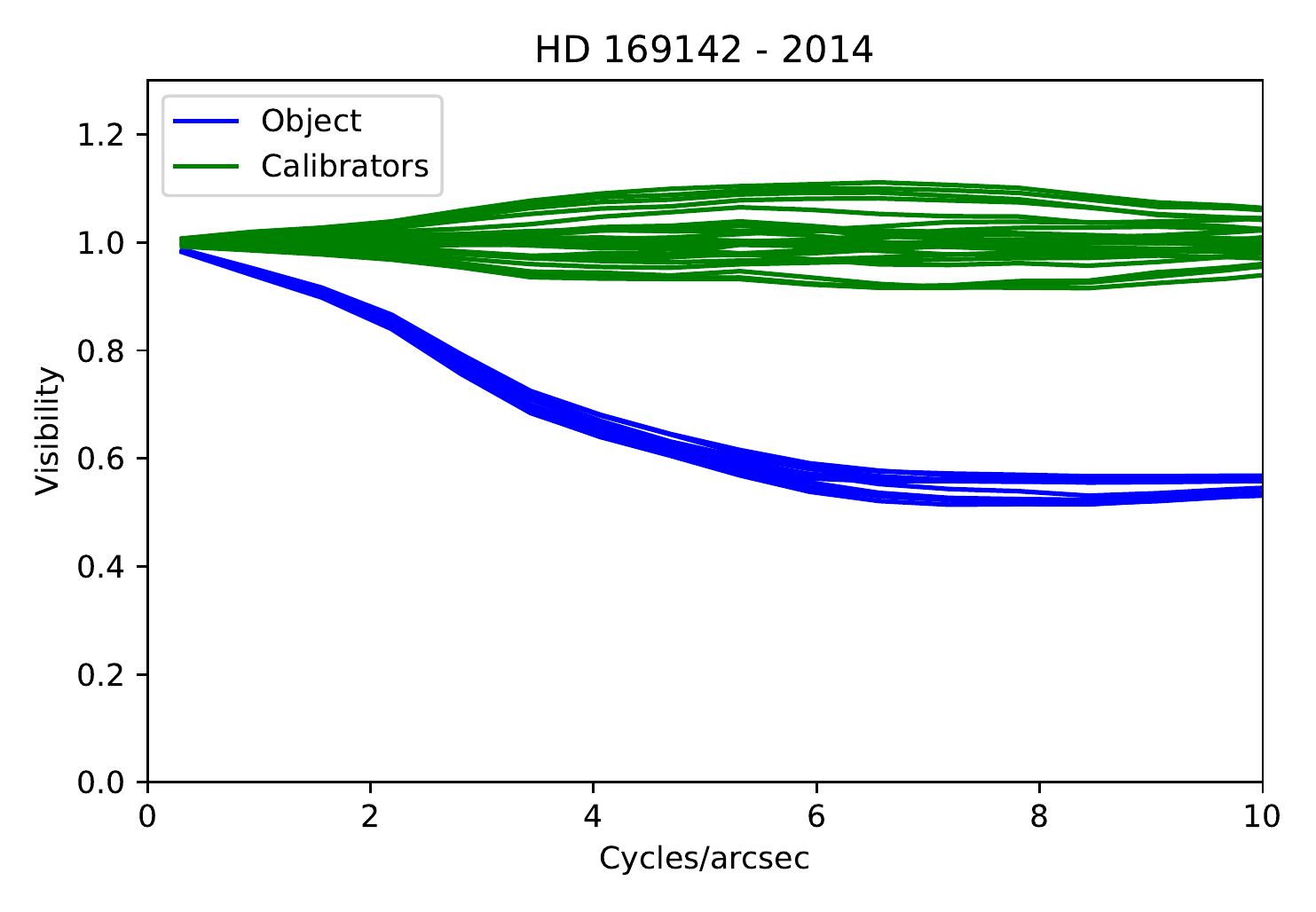}
                \caption{The same as Figure~\ref{figFourierIRS48}, for HD~169142 (for the 2014 epoch).}
  \label{figFourierHD169142}
\end{figure}

The presence of significant rings interior to the $\sim\,20\,\au$ wall inferred from previous data sets \citep[e.g.][for \irs~and \hd~respectively]{bruderer_gas_2014, osorio_imaging_2014} in the Richardson-Lucy deconvolved images is something that could in principle also arise from instability in the adaptive optics system. 
This possibility was ameliorated in our data collection by having multiple epochs on the targets and multiple calibrators at each epoch. 
In addition, Strehl ratios were reasonably high (0.5--0.7) and stable (within $\sim$10\% in the used images) given the long wavelength of our observation.

In order to quantitatively examine the effects of different calibrators, we performed a power spectrum analysis, as pioneered in speckle interferometry \citep{Labeyrie70}, where spatial power spectra were calculated for both individual calibrator stars and target stars, and then divided by the mean of the PSF calibrator power spectra in two dimensions. 
These calibrated power spectra are then equal to the power spectrum of the target itself, with the dispersion amongst the calibrator power spectra representing the uncertainty in this technique.

We plot the azimuthal averages of the visibilities (i.e. the square root of the power spectra) from each non-rejected frame of the target and calibrator in Figures~\ref{figFourierIRS48} and~\ref{figFourierHD169142} for the 2014 and 2015 epochs, divided by the mean azimuthally averaged visibilities. 
It can be seen that for both \hd~and \irs, the target visibility curves fall well outside the range of the calibrators, indicative of dominant structures that are fully resolved on 0.2--0.3" scales. 
The asymptotic visibilities of $\sim$0.26 for \irs~and $\sim$0.57 for \hd~measures directly that only $\sim$26\% and $\sim$57\% of the L' flux from \irs~and \hd~come from the sum of the central star and inner disc, in disagreement with previous models constrained by the SED \citep[e.g.][for \hd~and \irs~respectively]{seok_dust_2016, bruderer_gas_2014}.

\begin{figure}
   \centering
	\includegraphics[width=0.5\textwidth]{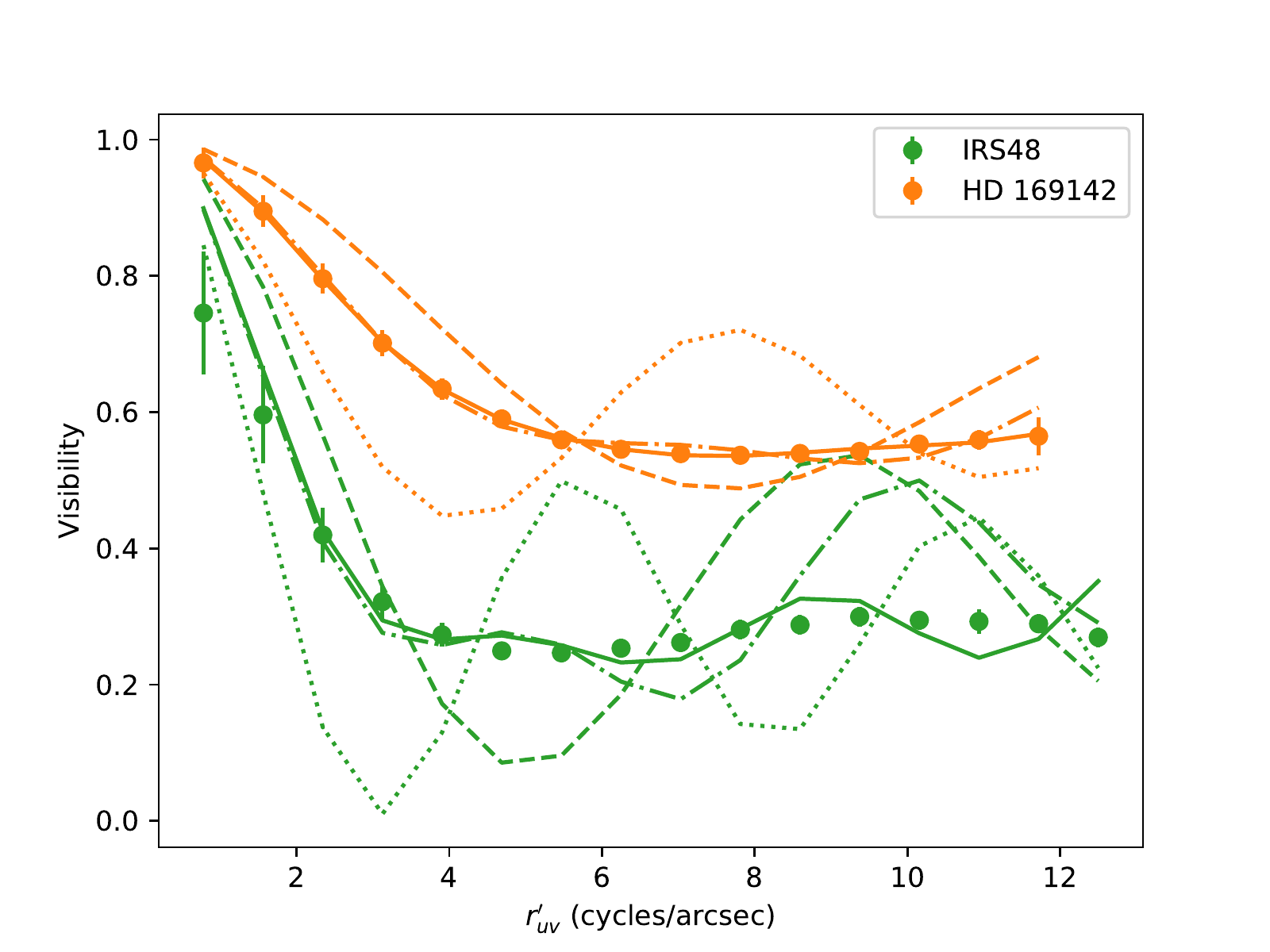}
	
                \caption{Deprojected, azimuthally averaged visibility curves for \irs~(green) and \hd~(orange). The circles represent the visibility measurements. Various models are fitted to these data: single rings at the previous literature location of walls (dotted lines); single rings at variable locations (dashed lines); two rings at variable locations (dot-dashed lines); and three rings at variable locations (solid lines). For both objects, one ring is a poor fit to this azimuthally averaged data, and at least 2 rings or a disc with this many free parameters, is required.}
  \label{vis_curve}
\end{figure}

In order to further analyse these visibilities, we used a model where each target was represented by an azimuthally symmetric inclined thin disc. The de-projected spatial frequency co-ordinate $r'_{uv}$ is derived from the $(u,v)$ coordinate using:
\begin{align}
r'_{uv} &= \sqrt{ r_{uv}^2 \cos(\theta_{uv} - \theta_{\rm maj})^2 + r_{uv}^2 \sin(\theta_{uv} - \theta_{\rm maj})^2 \cos(i) },
\end{align}
where $\theta_{uv}$ is the position angle of the $(u,v)$ coordinate, $r_{uv}$ its magnitude, $\theta_{\rm maj}$ the position angle of the disc major axis and $i$ the disc inclination. 
The Fourier power was then binned and averaged over this de-projected Fourier coordinate $r'_{uv}$ to enable robust data visualisation and azimuthally symmetric fitting. 
For \irs, an inclination of $55^\circ$ and a major axis position angle of $99^\circ$ were used, while for \hd, an inclination of $30^\circ$ and position angle of $13^\circ$ were used (based on literature values and our findings in Section \ref{sec:results} - where they differ from literature values). 
After azimuthally averaging the deprojected power spectra and taking the square root, we plot the visibility curves in Figure \ref{vis_curve}. 
Rising visibilities to mid or high spatial frequencies demonstrate that fits of ring-like or sharp-edged flux distributions are appropriate, rather than e.g. Gaussian or power-law dust distributions. 
We fitted several parameterized models to these distributions in order to determine the required model complexity (Table~\ref{tab:rad_flux}~and Figure~\ref{vis_curve}). 
One ring at the previous location of the wall inferred from the SED or long wavelength observations is clearly inconsistent with the data, which requires a more continuous distribution of emission, including an emission component inside the previous ($\sim20\,\au$) gap.

\begin{table}
\caption{Tested radii and fluxes for rings in \irs~and \hd, where $F$ refers to the fraction of flux that comes from each ring of radius $r$. The visibility curves generated by these model systems are shown in Figure \ref{vis_curve}}
\begin{center}
\footnotesize{{\begin{tabular}{cccccc}\hline
\multicolumn{6}{c}{\irs} \\
$F_1$ & $r_1\,(\au)$ & $F_2$ & $r_2\,(\au)$ & $F_3$ & $r_3\,(\au)$ \\
\hline
 0.71 & - & - & - & - & - \\
 0.66 &  14.3 & - & - & - & - \\
 0.38 &  12.7 &  0.29 &  25.2 & - & - \\
 0.22 &   9.8 &  0.26 &  17.0 &  0.20 &  26.9 \\
\hline
\hline
\multicolumn{6}{c}{\hd} \\
$F_1$ & $r_1\,(\au)$ & $F_2$ & $r_2\,(\au)$ & $F_3$ & $r_3\,(\au)$ \\
\hline
 0.40 & - & - & - & - & - \\
 0.37 &  11.0 & - & - & - & - \\
 0.26 &  10.7 &  0.14 &  20.7 & - & - \\
 0.24 &  10.2 &  0.12 &  18.6 &  0.03 &  28.4 \\
 \hline

\end{tabular}}}
\end{center}
\label{tab:rad_flux}
\end{table}


\section{Theoretical disc modelling} \label{sec:models}

In order to interpret the observational data, radiative transfer models of the discs were generated. 
We chose to follow previous literature on these discs, \citep[e.g.][\irs]{bruderer_gas_2014} and to fit power-law surface density distributions with multiple gaps. 
We recognise that alternative models with smoother distributions of dust and gas may fit our data equally well, and could be attempted as an extension to this work, constrained by hydrodynamical models or more detailed SED modelling, for example. 
Limitations of our work are discussed in Section \ref{sec:lims}.
The radiative transfer disc models used in this work are made using \radmc~\citep{dullemond_radmc-3d:_2012}\footnote{Code is available at: http://www.ita.uni-heidelberg.de/$\sim$dullemond/software/radmc-3d/}, which is a Monte Carlo radiative transfer code.
A forward modelling approach was taken; the models are generated by this code, rotated to match the observed position angle at each epoch, and convolved with a PSF so that a comparison can be made between the model and the observational data.

The radiative transfer models are discussed in Section \ref{sec:rt}, and the details of the parameters of the models are in Section \ref{sec:params}.

\subsection{Radiative transfer} \label{sec:rt}

The protoplanetary disc function of \radmcpy, the companion Python library to \radmc~\citep{dullemond_radmc-3d:_2012}, was used to create three dimensional disc density distributions with structures such as gaps and rings, which were then fed into \radmc.

Three model types are tested; a symmetric disc, an asymmetric disc and an asymmetric disc with a companion, with each model type being an extension on the previous one. These three model types are used to identify the presence of asymmetries in the data and attempt to characterise the asymmetry to see if it is point-like. 
To simulate an asymmetric disc, the central star can be moved slightly which gives an asymmetric dust illumination.
In some cases companions can cause the disc to become eccentric \citep{thalmann_imaging_2010}, meaning that the star is no longer at the centre of the disc. 
Sub-stellar companions are simulated by adding photospheric emission from a second point source to the \radmc~disc model.
The companion source is less massive, smaller in radius and cooler than the central star. 
The parameters involved in generating these models are discussed in Section \ref{sec:params}.

\subsection{Parameters} \label{sec:params}
\subsubsection{Globally constant parameters} \label{sec:both}

The stellar emission type, dust model, and stellar source type were held constant for both objects.
We use Kurucz\footnote{http://kurucz.harvard.edu/grids.html} \citep{castelli_new_2004} stellar models, with solar abundances and micro-turbulence of $2\,\mathrm{km\,s^{-1}}$ (an arbitrary choice, as we are not attempting to model the spectrum of the star in detail). The \radmc~emission source used for either object is a point source. 

The chosen dust model used for the final modelling of both objects is a fine-grained carbon and PAH dust mixture (henceforth CP dust).
The dust model chosen was the \citet{draine_infrared_2001, draine_infrared_2007} dust, which includes carbon in graphitic form and also grain sizes down to nm scale which are mostly PAHs.
The discs of both \irs~and \hd~are known to contain PAHs \citep[e.g.][\irs, both, \hd, respectively]{geers_spatially_2007, maaskant_polycyclic_2014, seok_dust_2016}. 
The reasoning behind the choice of dust is explained further in Section \ref{sec:dust}.
The dust best suited from the \citet{draine_infrared_2001, draine_infrared_2007} set of neutral dust was a grain-size of $5.6\times 10^{-3}$\,\micron~(5.6\,nm).
At 5.6\,nm the dust mixture can give strong PAH features, but also has a contribution from a more general graphitic source \citep{draine_infrared_2001, draine_interstellar_2003, draine_infrared_2007}.
This 5.6\,nm CP dust is used for the determination of the other disc parameters.

Other parameters that are the same for both objects include the number of photons (set to 10$^6$), and the spatial resolution. These were both set based on results of convergence tests, ensuring a reproducible output. 
The interested reader is directed to Appendix \ref{app:extra} for information on some of the other \radmc~parameters that were used and convergence studies.

\subsubsection{Object constant parameters} \label{sec:held}

The stellar temperature and mass are fixed for each object based on literature spectral type and age for each object, with the radius coarsely adjusted to match the observed visible flux (Table~\ref{tab:held}).
As we model the small-grained dust only, we can not fit directly to the gas density.
We choose to fix the mass of the disc (most of the mass is in the outer disc), to a value from the literature.
Instead of a varying disc mass, we have a dust-to-gas mass ratio that is independent of radius, with density (of dust and gas) that is able to deviate from dust-to-gas ratio at various radii, as discussed in the next section. 
Both of the objects studied in this work are known to have either outer rings or asymmetries observed in much longer wavelengths (e.g. mm wavelengths), so we include these known parameters as the outer wall of the disc, $r_{\mathrm{wall}}$ and outer disc dust depletion $\delta_{\mathrm{wall}}$.
The flux ratio refers to the ratio of flux between the star and the disc, and is discussed in more detail in Sections \ref{sec:priors} and \ref{sec:dust}.

\begin{table}
\caption{Fixed parameters for \irs~and \hd.}
\begin{center}
\footnotesize{{\begin{tabular}{ccc}\hline
\multicolumn{3}{c}{\irs} \\
Set Parameter & Value & Reference \\
\hline
Stellar Type & A0 & 1, 2 \\
Star Temperature & $9000\,\K$ & 2 \\
Star Mass & $2.0\,\msolar$ & 2 \\
Star Radius & $2.24\,\rsol$ & Assumed \\
 &  &  (young A0 star) \\
Disc Mass & $10^{-4}\,\msolar$ & 3 \\
$r_{\mathrm{wall}}$  & $67\,\au$ & 4, adjusted to new \\
 & & distance \\
$\delta_{\mathrm{wall}}$ & 0.1 & 3 \\
Dust Type & 5.6\,nm C \& PAH & Assumed, 5  \\
Distance to object & $134.4\,\pc$ & 6 \\
Flux Ratio & 7.8 & This work \\
\hline
\hline
\multicolumn{3}{c}{\hd} \\
Set Parameter & Value & Reference \\
\hline
Stellar Type & A5 & 7 \\
Star Temperature & $8250\,\K$ & 7 \\
Star Mass & $1.65\,\msolar$ & 7 \\
Star Radius & $1.56\,\rsol$ & Assumed, 7 \\
Disc Mass & $10^{-3}\,\msolar$ & Assumed, 7 \\
$r_{\mathrm{wall}}$ & $55\,\au$ & 8, 9\\
$\delta_{\mathrm{wall}}$ & 0.1 & Assumed \\
Dust Type & 5.6\,nm C \& PAH & Assumed, 5 \\
Distance to object & $114\,\pc$ & 6 \\
Flux Ratio & 5.2 & This work\\
 \hline

\end{tabular}}}
\begin{minipage}{\linewidth}
\vspace{0.1cm}
\footnotesize{\textbf{Notes.} 1. \citet{mcclure_evolutionary_2010}; 2. \citet{brown_30_2012}; 3. \citet{bruderer_gas_2014}; 4. \citet{van_der_marel_major_2013}; 5. \citet{maaskant_polycyclic_2014}; 6. \citet{gaia_collaboration_gaia_2018}; 7. \citet{seok_dust_2016}; 8. \citet{quanz_gaps_2013}; 9. \citet{osorio_imaging_2014}.} 
\end{minipage}
\end{center}
\label{tab:held}
\end{table}

\subsubsection{Parameters varied for each object} \label{sec:varied}

\begin{figure}
   \centering
	\includegraphics[width=0.5\textwidth]{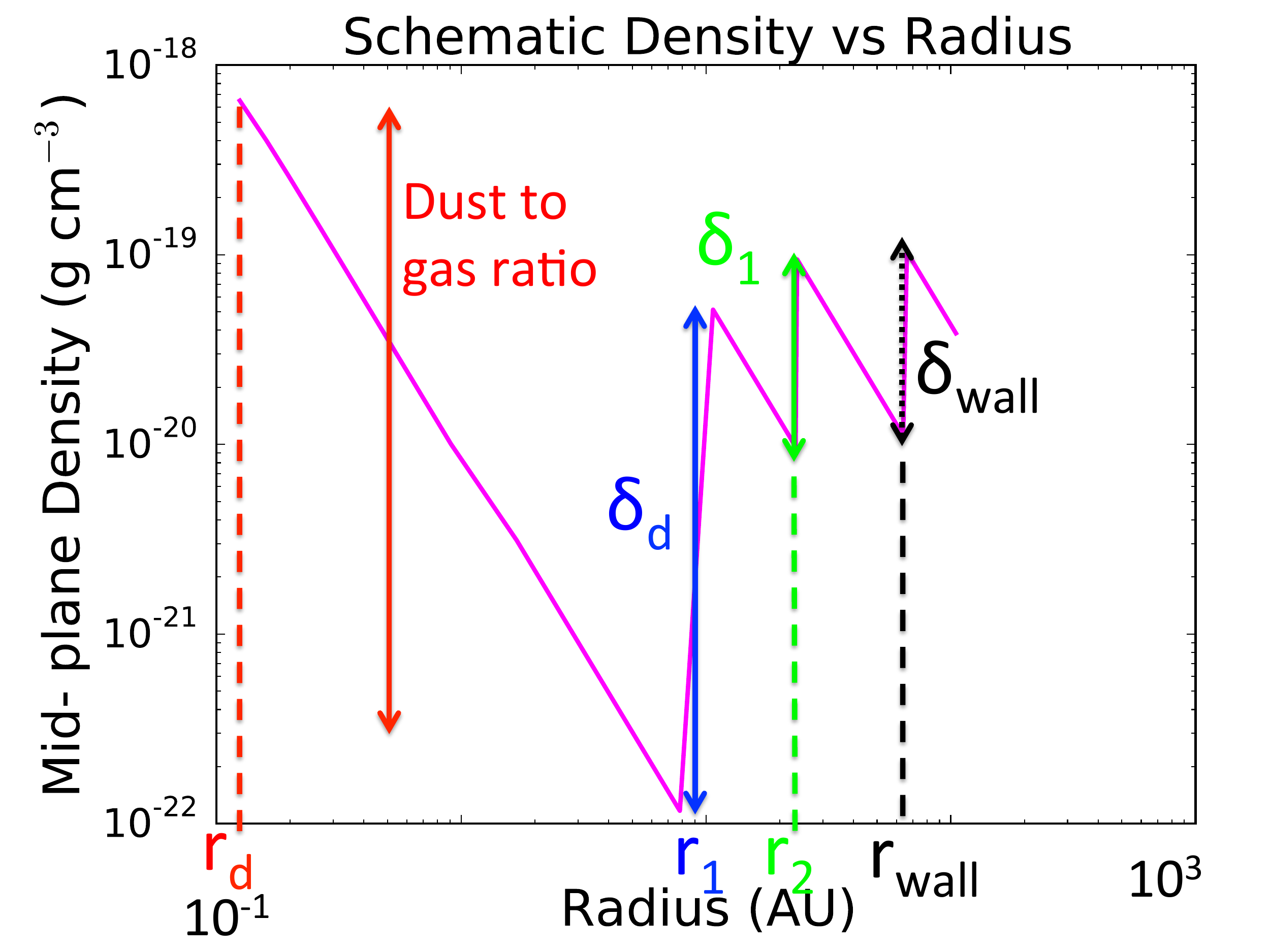}
                \caption{Shown here in magenta is a plot of the typical mid-plane density structure as a function of radius in the disc model. The parameters relating to the radii and density are marked. The dust to gas ratio, represented by the red arrow, moves the whole profile (magenta line) up or down. The first depletion factor, $\delta_{d}$ is that of the inner disc, and is marked with a blue arrow. It determines how depleted the inner disc is of material. The inner disc extends from $r_{\mathrm{d}}$, marked in red, to $r_{\mathrm{1}}$, marked in blue. The green arrow represents the depletion factor $\delta_{1}$ between the two walls, $r_{\mathrm{1}}$ (blue) and $r_{\mathrm{2}}$ (green) that are allowed to vary in radius. The black arrow shows the set depletion factor, $\delta_{\mathrm{wall}}$ for between these inner regions and the rest of the disc, at a distance of $r_{\mathrm{wall}}$ (black).}
  \label{density_change}
\end{figure}

\begin{figure*}
   \centering
	\includegraphics[width=1.\textwidth]{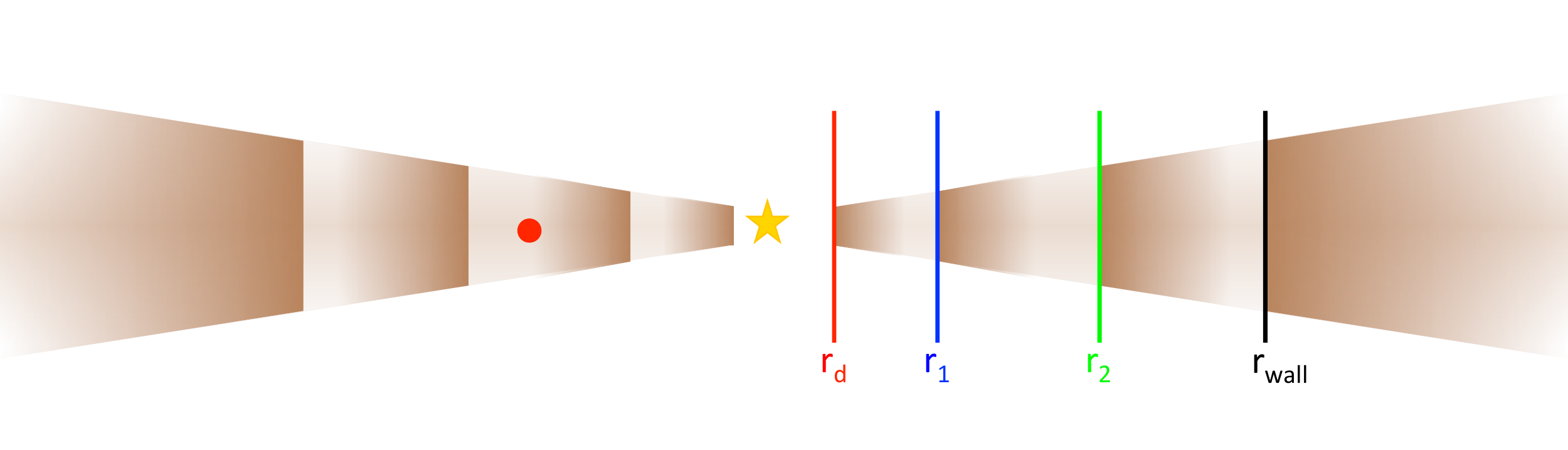}
                \caption{This is a schematic of our disc model. The colours are the same as in Figure \ref{density_change} - the red line indicates the inner edge of the disc, at the dust sublimation radius ($r_{\mathrm{d}}$), and the black line is the radius of the wall known from mm-wavelength observations ($r_{\mathrm{wall}}$). The blue line marks the radius of our new wall ($r_{\mathrm{1}}$), and the green line marks the radius of the wall that is fitted based on mid-IR observations ($r_{\mathrm{2}}$). The gradient in the brown colour indicates the density structure, which is illustrated in Figure \ref{density_change}. The star is off-centre, to demonstrate how the asymmetry is introduced into the simulation, and the red circle indicates a companion, which can be placed anywhere in the midplane of the disc.}
  \label{disc_structure}
\end{figure*}

The symmetric model uses the following eight parameters:  the overall dust-to-gas ratio; the inner radius of the disc $r_{d}$; the depletion factor of the inner disc, $\delta_{d}$; the radius of the first wall, $r_{1}$; the depletion of the first gap, $\delta_{1}$; the radius of the second wall, $r_{2}$; the inclination of the disc; and the position angle of the object on the sky. 
The first six parameters are illustrated in a schematic plot of the midplane density of a disc in Figure \ref{density_change}, and a schematic indicating the radii and other parameters is shown in Figure \ref{disc_structure}.
The inclination is whether the object is seen face-on, edge-on or somewhere in-between.
The position angle quoted is the position angle of the disc major axis from North towards East, with the near side of the modelled disc at this position angle plus 90 degrees. 
Note, however, that it is uncertain which minor axis corresponds to the near side.

The use of two walls in the disc at $r_{1}$ and $r_{2}$ allows us to explore the position of the known wall at $r_{2}$, while introducing a new wall at a closer radius, $r_{1}$. 
Both the objects investigated here are known to have rings of emission at a radius of $\sim20\,\au$, and so to recover the position of that ring, while also fitting our data we use two radii here.

The asymmetric model case moves the central star. 
Moving the central star simulates the appearance of an asymmetric disc, without making an asymmetric dust distribution, which could add an arbitrary degree of complexity.
For YSOs, it is possible for the star to be slightly offset from the centre of the disc, for example this was one explanation for the peri-centre offset of LkCa15 \citep[in e.g.,][]{thalmann_imaging_2010}.
The eccentricity of the disc in systems where the star has a peri-centre offset is expected to be driven by the presence of a companion.
The off-centre star placement means that one side of the disc is more strongly illuminated than the other, and is indicated by the position of the star in Figure \ref{disc_structure}.
The asymmetry parameters come from changing the $x$ and $y$ positions of the central star (units of this movement are in $\au$).
The positive $x$ direction is to the right along the semi-major axis, and the positive $y$ direction is to the right along the semi-minor axis.
These parameters are converted to a separation in mas and position angle offset when presented in Section \ref{sec:results}.

The companion model is constructed by adding a second point source to the asymmetric disc model. 
We do not explicitly suggest that a point-like asymmetry is likely to be an accreting exoplanet at our poor angular resolution, but include a point source for simplicity of parametric modelling.
The parameters that are varied in the model for the companion to ensure a better fit are the radius and $x$ and $y$ positions of the companion (where the $x$ and $y$ directions are the same as for the asymmetry).
As with the star parameters, the position of the companion is converted to a separation given in mas and a position angle, so that it is more easily compared to literature values for the position of potential companions. 
The radius of the planet was used as a proxy for the brightness within our simulation setup and it is converted to a contrast of the companion to the total model, given in magnitudes.
Please see Appendix \ref{app:rad} for details on the companion parameters. 
An example of a companion is denoted by the red circle in Figure \ref{disc_structure}.

\subsubsection{Priors} \label{sec:priors}

For some of the parameters that will be explored, Bayesian priors were set.
The priors for the radii make sure that they stay in the order of $r_{d} < r_{1} < r_{2} < r_{\rm wall}$. 
As minimum inner disc radius, we set 0.1 AU, however, this prior is not important in the determination of the likelihood peak.

When the star is placed off-centre, there is a prior on the displacement so that the displacement does not exceed a radius of $1\,\au$ from its central position, which we find does not affect the likelihood peak.

There is also a prior on the flux ratio of the star to the total flux (star+disc).
The flux ratio prior is used so that without actually modelling the full SED (which would require multiple dust models), 
we can still constrain the amount of dust in the unresolved inner disc, and to have a self-consistent disc model that fully takes 
into account shadowing of the outer disc by the inner disc.

For both \irs~and \hd, we fit the flux ratio of the entire disc and star system to the star.
We calculate the flux ratio to fit by taking the stellar photosphere models of the objects from our simulations and literature observations at 3.4 (WISE) or 3.6\micron~(Spitzer/IRAC) photometry.
To find our model flux ratio, we calculate the total intensity of a model with just a star, and for a model of a star and a disc,  and then calculate the ratio of these intensities.

For \irs~the flux ratio used was $\sim7.8$ and for \hd~it was $\sim5.2$.
Uncertainties for each of the flux ratios are estimated for use in the \emcee~part of the method.
The flux ratio places a constraint on the brightness of the disc in the model, and in particular, the unresolved inner disc.

\subsubsection{Limitations}
\label{sec:lims}

There are some limitations to our method and the way our models are constructed.
Our models have a similar power-law density structure to those used in other works \citep[e.g.][\irs, both, respectively]{bruderer_gas_2014, maaskant_polycyclic_2014}, as seen in Figure \ref{density_change}.
Alternative density structures (e.g. without discontinuities) would also fit the data, however we choose to analyse one class of models with sufficient parameters.
We consider a phenomenological model, and do not test whether the vertical structure is in equilibrium. 
We also assume that the distances to the objects are fixed, and do not use the uncertainties associated with them in our modelling.
The interested reader is directed to Appendix \ref{app:extra} for more details on the parameters not discussed in Section \ref{sec:params}.

\subsection{Convolving the models} \label{sec:conv}

The final step before the model image can be compared to the observational data is to convolve it with a PSF, chosen from our PSF library generated by the observation of calibrator objects.
The model image is rotated to the same position angle as the observational data, then convolved with each of the PSFs in the relevant PSF library, shifted on a sub-pixel scale to match the position of the target data, and then the best PSF for convolution is chosen for each model image.
The best PSF is found for each of the rotated images by comparing with observational data and calculating the \chisquare~for each of the models, and the best convolved model is the one with the lowest \chisquare.  

The process of choosing the best convolved image is described by Equation \ref{eqn:psf},
\begin{equation} \label{eqn:psf}
\chi^2_{\rm{shot,} i} = \mathrm{min}_{j}\sum_k \frac{(D_{i, k} - (M\otimes P_{j})_k)^2}{\sigma_k^2}.
\end{equation}
The \chisquare~for each data image $D$, of which there are  $i$, is calculated by finding the minimum of \chisquare~value over the set of PSFs $P$, of which there are  $j$, over all of the pixels with the index $k$; where $\sigma_k^2$ is the variance over those pixels, accounting for readout noise, target shot noise and background shot noise. Note that we deliberately choose not to attempt to model speckles and adaptive optics phase noise as additional uncertainties, as it is in principle not a fundamental noise source in a well characterised and calibrated adaptive optics imager. We do, however, scale our final reported uncertainties appropriately, as described in Section~\ref{sec:fits}. 

The rotated model image is convolved with all of the PSFs, and the best of these is chosen to calculate the \chisquare~for the data image $D_i$.

These best \chisquare~values for each image are then summed to find the total \chisquare~value, $\chi^2_\mathrm{tot}$,  
\begin{equation} \label{eqn:chitot}
\chi^2_\mathrm{tot} = \sum_i \chi^2_{\mathrm{shot}, i} .
\end{equation}

This process of choosing the single best calibrator can in principle under-fit the data
in a similar way that the locally optimized combination of images (LOCI) algorithm under-estimates the brightness of detected 
companions \citep{Soummer12}. We checked that this type of systematic was not severe by 
verifying that many different calibrator images were chosen for differing target images. 
Given how flat the squared visibility versus baseline curve of the calibrators are
when calibrated against each other (see Figures~\ref{figFourierIRS48} and ~\ref{figFourierHD169142}), this systematic is limited to
missing flux in our model at up to the few percent level in the outer portion of the disc.
A more robust algorithm would need to develop a significantly larger set of model PSFs
(e.g. from a linear combination of calibrator images) and marginalise over all possible
PSFs, as described in \citet{Ireland16}.

\subsection{Model parameter estimation via MC in MC} \label{sec:mcinmc}

All of the aforementioned steps and the parameter estimation are completed using our MCinMC technique.
MCinMC is the use of the Monte Carlo radiative transfer code \radmc~in combination with the Markov Chain Monte Carlo (MCMC) code \emcee~\citep{foreman-mackey_emcee:_2013}. 
Given the nature of the large parameter space, and the risk that some of the parameters may be correlated, an MCMC approach was taken to ensure that the best-fit parameters were the global minimum, rather than a local minimum.
\emcee~is an affine-invariant ensemble sampler, meaning that it is not affected by covariances between parameters.
\emcee~uses a number of walkers to explore a parameter space. 
In this work the walkers are initialised from a group of points in the N-dimensional parameter space and the parameter space is explored.
\emcee~can be easily made parallel onto different compute cores to make the computations more efficient.

Once the best \chisquare~for each individual data image and its corresponding rotated, convolved model image is calculated, the \chisquare~values are summed and used to calculate the log likelihood of the model;
\begin{equation} \label{eqn:lnlike}
ln(\mathrm{likelihood}) =  \frac{-\chi^2_\mathrm{tot}}{2T_{\mathrm{MC}}},
\end{equation}
where the total \chisquare~is that calculated in Equation \ref{eqn:chitot}, and $T_{\mathrm{MC}}$ is the \emcee~temperature, set so that scaling the error bars by $\sqrt{T_{\mathrm{MC}}}$ would give a reduced $\chi^2$ of 1.  
We choose this approach rather than adding additional uncertainty estimates to our data variance, in order to show how much better the fits could be if PSF models were improved.
\emcee~uses the log likelihood to find the best model for the data.
The reduced \chisquare~ is calculated from this log likelihood, given by:
\begin{equation} \label{eqn:redchi}
\chi^2 =  \frac{-ln(\mathrm{likelihood})\times2\times T_{\mathrm{MC}}}{n_{\mathrm{pixels}}\times n_{\mathrm{images}}},
\end{equation}
where the number of pixels in the image $n_{\mathrm{pixels}}$, and the number of images $n_{\mathrm{images}}$ normalise over the degrees of freedom, making this the reduced \chisquare.
The number of pixels in the image ($n_{\mathrm{pixels}}$), is actually the number used in the calculation, rather than the total number of pixels in the image.
The number of images ($n_{\mathrm{images}}$) is the number of target star images.


\section{Synthetic Data Set}
\label{sec:synth}

Our method was tested by generating a synthetic data set within MCinMC and then running a full parameter exploration on the synthetic data.
The synthetic data set was generated by convolving a \radmc~model, with parameters as listed in Table \ref{tab:synth_set}, and the values in the column labelled `Target' in Table \ref{tab:synth_mc_res} with a sequence of calibrator (PSF) images from HD 167666.
When this data set was tested a different sequence of calibrator images (HD 170768) were used for generating the convolved model for comparison with the data.
It was found that the method could recover the parameters of the initial model to within 5 standard deviations for all parameters, with our approximate method of accounting for point-spread function fitting uncertainties as described in Section~\ref{sec:fits}.

\begin{table}
\caption{General parameters for the synthetic data set.} 
\label{tab:synth_set}
\begin{center}
{\footnotesize\begin{tabular}{cc}\hline
\multicolumn{2}{c}{Parameters that will remain set.} \\
Parameter & Value  \\
\hline 
Star Temperature & $10000\,\K$  \\
Star Mass & $2.0\,\msolar$  \\
Star Radius & $2.0\,\rsol$  \\
Disc Mass & $10^{-3}\,\msolar$  \\
$r_{\mathrm{wall}}$ & $50\,\au$  \\
$\delta_{\mathrm{wall}}$ & 0.1  \\
Dust Type & 5.6\,nm~CP dust   \\
Distance to object & $100\,\pc$  \\
\hline
\end{tabular}}
\end{center}
\end{table}

\subsection{Testing the Synthetic Data Set with MCinMC}

\begin{figure*}
\centering
\includegraphics[width=1.0\textwidth]{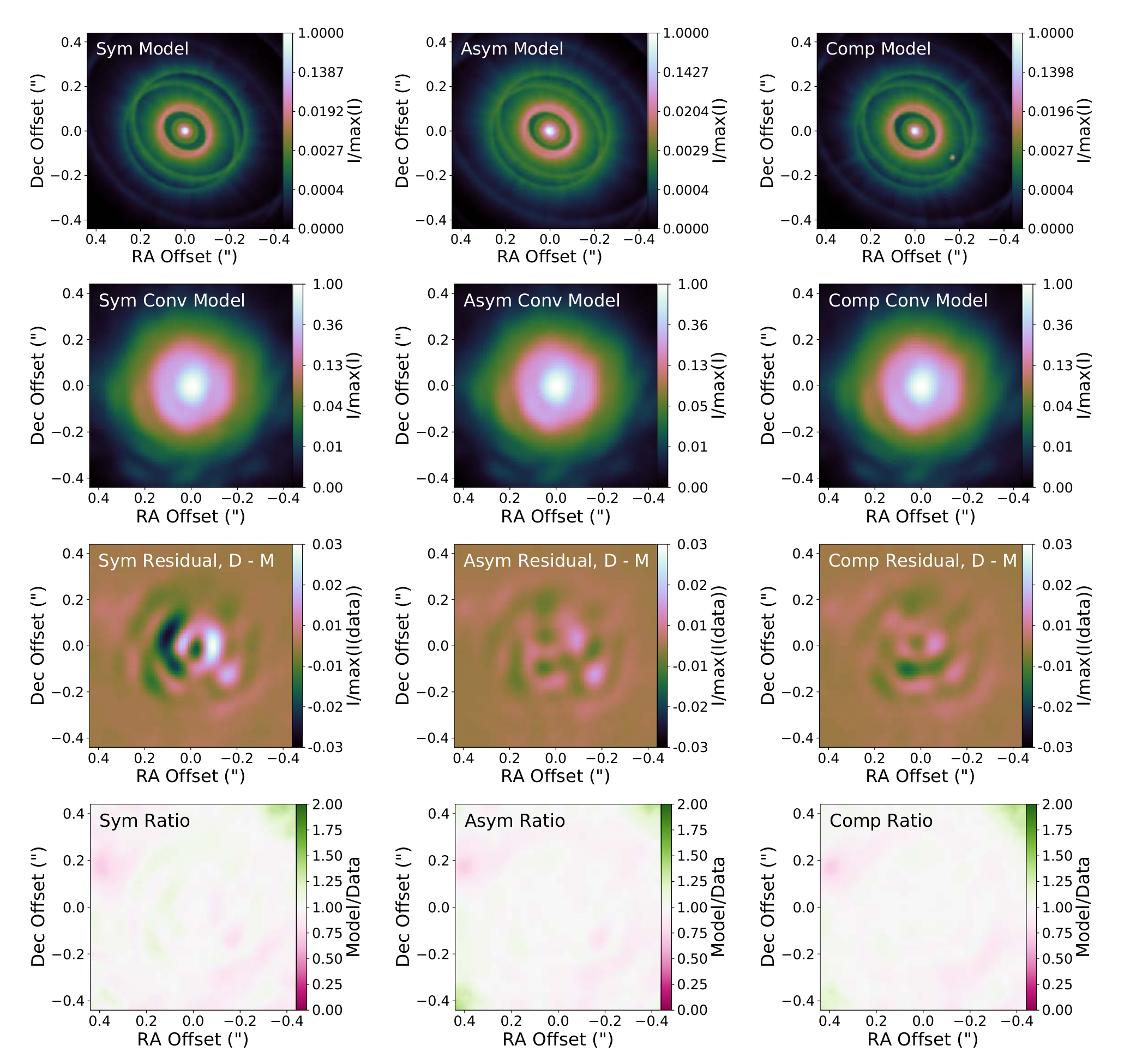}
  \caption{Converged results from the modelling of the synthetic companion data set. The left column shows the symmetric model results, the middle shows the asymmetric model results, and the right shows the asymmetric with companion model results. The top row is the model image for each model type. The second row is the convolved model. The third row is the residual of the observed target data image minus the convolved model image. The fourth row is the ratio of the convolved model divided by the observed target data image.}
  \label{synth_fig_results}
\end{figure*}

A model with a disc asymmetry and companion was used to generate the synthetic data set and then this data set was tested in the MCinMC code.
The data set was tested as a symmetric, asymmetric and asymmetric with companion model.
The results of this exploration are shown in Table \ref{tab:synth_mc_res}, with figures of the results in Figure \ref{synth_fig_results}.
The target parameters are those used to construct the data set, listed as Target in Table \ref{tab:synth_mc_res}.

\begin{table*}
\caption{Results for the synthetic companion data.} \label{tab:synth_mc_res}
\begin{center}
{\footnotesize\begin{tabular}{ccccc}\hline

\multicolumn{5}{c}{Synthetic Companion Data} \\
Parameter & Target & Symmetric Model & Asymmetric Model  & Companion Model \\
\hline 
Dust to gas ratio           & $1\times10^{-2}$ & 1.51${\pm0.47}\times 10^{-2}$ & 6.56${\pm1.80}\times 10^{-3}$  & 1.12${\pm0.11}\times 10^{-2}$\\
$\delta_{d}$ & $1\times10^{-4}$ & 7.87${\pm1.61}\times 10^{-5}$ & 1.27${\pm0.20}\times 10^{-4}$  & 9.83${\pm0.60}\times 10^{-5}$\\
$\delta_{1}$       & $1\times10^{-2}$ & 8.89${\pm1.42}\times 10^{-3}$ & 1.22${\pm0.14}\times 10^{-2}$  & 1.11${\pm0.08}\times 10^{-2}$\\
$r_{d}$ $(\au)$   & 0.5 & 0.39${\pm0.07}$ & 0.64${\pm0.08}$  & 0.42${\pm0.03}$\\
$r_{d}$ $(\rm{mas})$   & 5 & 3.9$\pm{0.7}$ & 6.4$\pm{0.8}$ & 4.2$\pm{0.3}$  \\
$r_{1}$ $(\au)$        & 10.0 & 10.0${\pm0.1}$                & 10.2${\pm0.1}$                 & 9.95${\pm0.10}$\\
$r_{1}$ $(\rm{mas})$ & 100 & 100${\pm1}$ & 102${\pm1}$ & 99.5${\pm1.0}$ \\
$r_{2}$ $(\au)$         & 25.0 & 23.9${\pm0.6}$                & 23.3${\pm0.4}$                 & 23.6${\pm0.5}$\\
$r_{2}$ $(\rm{mas})$ & 250 & 239${\pm6}$ & 233${\pm4}$ & 236${\pm5}$ \\
Inclination $(\degree)$     & 30.0 & 29.7${\pm1.2}$                & 31.9${\pm1.1}$                 & 30.7${\pm1.2}$\\
Position angle $(\degree)$  & 50.0 & 54.1${\pm2.9}$                & 57.4${\pm2.2}$                 & 56.4${\pm2.0}$\\
Star offset $(\rm{mas})$ & 1.3  & - & 1.6$\pm{0.3}$ & 1.1$\pm{0.1}$ \\
Star position angle $(\degree)$ & 90.9 & - & 79$\pm{15}$ & 75$\pm{13}$ \\
Companion offset $(\rm{mas})$ & 180 & - & - &  186$\pm{2}$\\
Companion position angle $(\degree)$ & 232.8 & - & - & 235.6$\pm{2.2}$ \\
Companion contrast (mag) & 5.4 & - & - & 5.7$\pm{0.2}$ \\
Flux Ratio          & 30.31 & 30.33 & 30.32 & 30.32  \\
\hline
reduced \chisquare & - & 67.5 & 27.3 & 24.4 \\
\hline
\end{tabular}}
\end{center}
\end{table*}

The symmetric model is the worst fit for the synthetic data, at a reduced \chisquare~of 67.5.
For the symmetric model only the inner wall radius and the inclination are the same (within uncertainties) as those used to generate the data, however most were still within a few standard deviations of the true values.
Since the data is asymmetric with a companion, it is not surprising that the symmetric model does not fit the data well.
The \chisquare~for the symmetric model is more than twice that of the asymmetric model.

Most parameters for the asymmetric model are closer to those used to generate the data than the symmetric results are, and those closest to the parameters used are the two depletion factors, the radius of the inner disc, the radius of the first wall, the inclination, and the asymmetry parameters. 
The asymmetric model \chisquare~of 27.3 is a large improvement over the symmetric fit.

The companion model has an improved, but similar \chisquare~to the asymmetric model. 
The resulting fits to the 8--13 parameters for the companion fit are mostly within three standard deviations of the data. 
When the model is convolved with the PSF to generate the synthetic data, the partly-resolved companion gives a signal similar to a low-order aberration, and our simplified error model does not take this into account adequately. Note that in this case, the companion separation is dominated by the $x$ position (causing the slight deviations in the companion offset and position angle), which agrees reasonably well with the input model.

The residual images for both the asymmetric and companion models have obviously lower residuals than the symmetric model.
The symmetric model residual shows that there is an asymmetry in the data that is not being fit with this model.
The asymmetric and companion model residuals mostly show the noise in the PSFs, although there is a slight difference between the two at the companion position, showing that the companion model fits a feature where the asymmetric model cannot.
The small uncertainty on the brightness of the companion and its position indicate that there is a strong asymmetry in that part of the disc.

\subsection{Summary of Synthetic Data Set Results}
In this test our method is able to distinguish between a symmetric or asymmetric disc, but not as well between an asymmetry caused by an eccentric disc or a point-like asymmetry.
The symmetric model will not fit well to a disc with obvious asymmetries.
It is difficult to determine the cause of an asymmetry with our method, however we are sensitive to whether an asymmetry is present, and the strength of the asymmetry.

Our method is limited by the uncertainties in the PSFs, making it difficult to determine the difference between a disc asymmetry and a point-like asymmetry in the disc.
Given the faint nature of planetary companions it can be difficult to determine the difference between noise, disc asymmetries and a point-like asymmetry.

\section{Results} \label{sec:results}

\subsection{Dust} \label{sec:dust}
\begin{figure}
        \centering
	\includegraphics[width=0.5\textwidth]{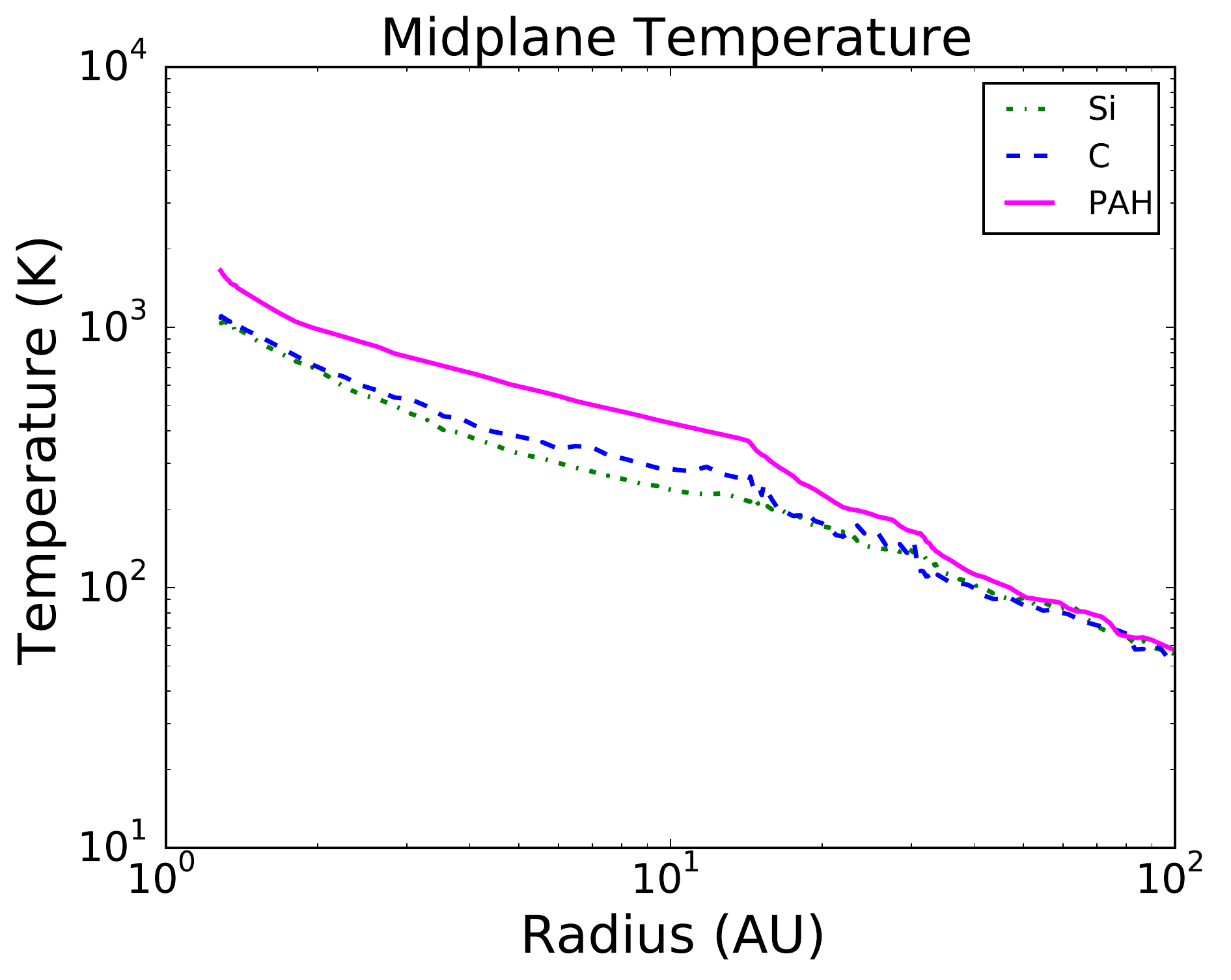}
    \caption{Temperature plots in the disc midplane for the different types of dust. Magenta solid line is CP dust, blue dashed line is carbon dust and the green dot-dashed line is silicate dust. The CP dust is warmer than the other dust types, particularly in the inner regions of the disc.}
    \label{dust_temp}
\end{figure}

It was initially found that carbon dust gave a better result (lower reduced \chisquare) than silicate dust; and that dust containing carbon and PAHs gave a better result than plain carbon dust.
Carbon produced better models for \irs~than silicate dust could, because silicates were too cool to produce the emission at $\sim$10\,AU radii (Figure~\ref{dust_temp}).
Relative to the absorption of stellar photons at $\la1$\,$\mu$m wavelengths, silicate dust has a large emission feature at 11\,\micron~which effectively cools the dust \citep{draine_interstellar_2003}.

Using carbon dust, it was not possible to simultaneously have the $\sim$13\,AU ring bright enough to fit our adaptive optics data and still have an inner disc bright enough to fit the SED.
The amount of carbon dust needed in the inner disc to fit the SED would cause too much shadowing on the 13\,AU ring. 
Therefore, a dust which could more effectively absorb blue and ultraviolet wavelengths was needed.

Both of the discs studied in this work have clear PAH features in their SEDs \citep[e.g.][\irs, \hd, both, respectively]{bruderer_gas_2014, seok_dust_2016, maaskant_polycyclic_2014}.
In our attempts to fit the near-IR excesses seen in the SED, CP dust was used, and the same dust was used for both objects due to the similarities in their SEDs.
Given that we are not trying to fit the whole SED, we chose a dust model from the \citet{draine_infrared_2001, draine_infrared_2007} dust, which includes fine grained graphitic carbon, and PAHs.
From inspection of the SEDs the dust chosen from the \citet{draine_infrared_2001, draine_infrared_2007} set of neutral CP dust has a grain-size of 5.6\,nm.
Even though it is known that some of the PAHs in these discs are ionised \citep{maaskant_polycyclic_2014}, we chose the neutral dust to use in both cases because it was similar to the PAH emission of the objects and able to replicate the near-IR excess.
For an investigation of how differently sized dust grains from the \citet{draine_infrared_2001, draine_infrared_2007} set of neutral CP dust change the fit to the SED, see the Appendix (\ref{sec:seds_dust}).

The reason that the smaller grains (CP dust) better recover the near-IR excess is that they are generally warmer than larger grains (see Figure \ref{dust_temp}), due to their
relatively higher UV absorption.
There is also no need to invoke quantum heating in this case, as $\sim5$\,nm radius dust has a sufficient number of atoms to not be significantly warmed by single photon absorption events. 
Thus quantum heating in the disc is ignored here, though it may become required to model images at all wavelengths (especially even shorter wavelengths) simultaneously.

One consideration to be able to have both a bright inner disc and bright walls at 13 and $30\,\au$ (as is needed for \irs, and similar to what is needed in \hd) would be to have in inclined inner disc.
With an inclined inner disc, carbon dust could still be used and the disc walls would still be bright.
However, for both \irs~and \hd, the deconvolved images show symmetric features, with no evidence for the shadows that would be expected at opposite sides of the disc if the inner disc was inclined.
Therefore an inclined inner disc with carbon dust is not considered here and we instead use CP dust.

A mix of different dust types would benefit the modelling of the disc and the SED simultaneously.
More freedom in the dust distribution and composition would allow for better modelling of disc features, and possibly determine the difference between a disc feature or a companion.
This was not justified for our current method, given the high speckle residuals, which were at the same level for our data sets as for synthetic data sets using multiple 
calibrator stars.

\subsection{Fits to \irs~and \hd} \label{sec:fits}

The best-fit model parameters for both objects are shown in Table \ref{tab:results_mean}. 
The primary fitting for \hd~was done for the 2014 data and for \irs~it was done for the 2015 data, the fit for the best fit model for each of these data sets is then tested on the 2016 data. 
We do not report independent fits to the 2016 data because as mentioned in section \ref{sec:datred}, some observations were taken through cloud.

In the following sections the results of each model type for each object will be discussed.
For each object there was a symmetric, an asymmetric and a companion model tested. 
The model setups are described in Section \ref{sec:params}.
Figures \ref{irs48_models} and \ref{hd_models} show each of these models; their convolved counterpart; the residual of the data and the model; and the ratio of the data and the model, for \irs~and \hd~respectively, and are discussed in Sections \ref{res_irs}, and \ref{res_hd}.

The values and uncertainties quoted in Table \ref{tab:results_mean} are the mean and standard deviation of the results found with \emcee.
The data used in this work has uncertainties with several components, including background and speckle noise.

We choose to only consider the target and background shot noise to give the reader a clear indication on the scope of improvement in model fitting that would be possible with an improved point-spread function model (e.g., interpolation between PSFs).
However, we also need to robustly estimate statistical uncertainties from our current fitting method, and choose to do this by setting the temperature in the Monte-Carlo Markov Chain.

\begin{table*}
\caption{Best fit model parameters for 2015 \irs~data and 2014 \hd~data, using mean and standard deviation. Best-fit models were not calculated for the 2016 data of both objects, however the best-fit models from the other years were tested on the 2016 data.} \label{tab:results_mean}
\begin{center}
{\footnotesize\begin{tabular}{cccc}\hline
\multicolumn{4}{c}{\irs} \\
Parameter & Symmetric & Asymmetric & Companion \\
\hline 
Dust to gas ratio          &3.10${\pm0.66}\times 10^{-3}$&3.20${\pm0.26}\times 10^{-3}$ &2.06${\pm0.09}\times 10^{-3}$\\
$\delta_{d}$ &2.48${\pm0.43}\times 10^{-4}$&2.11${\pm0.23}\times 10^{-4}$ &4.20${\pm0.35}\times 10^{-4}$\\
$\delta_{1}$       &1.01${\pm0.10}\times 10^{-2}$&9.40${\pm0.61}\times 10^{-3}$ &1.24${\pm0.06}\times 10^{-2}$\\
$r_{d}$ $(\au)$  &1.6${\pm0.2}$               &1.4${\pm0.1}$                &2.0${\pm0.1}$\\
$r_{d}$ $(\rm{mas})$& $12{\pm1}$ & $10{\pm1}$ & $15{\pm1}$ \\
$r_{1}$ $(\au)$       &15.0${\pm0.3}$               &14.6${\pm0.1}$                &15.2${\pm0.2}$\\
$r_{1}$ $(\rm{mas})$& $112{\pm2}$ & $108{\pm1}$ & $113{\pm1}$\\
$r_{2}$ $(\au)$        &32.1${\pm0.9}$               &31.0${\pm0.3}$                &31.3${\pm0.3}$\\
$r_{2}$ $(\rm{mas})$& $238{\pm7}$ & $231{\pm2}$ & $233{\pm2}$ \\
Inclination $(\degree)$    &54.8${\pm0.5}$               &54.2${\pm0.3}$                &52.4${\pm0.3}$\\
Position angle $(\degree)$ &99.0${\pm0.6}$               &98.8${\pm0.3}$                &96.5${\pm0.4}$\\
Star offset ($\rm{mas}$) & - & 5.0$\pm0.4$  & 5.3$\pm0.3$ \\
Star offset position angle $(\degree)$ & - & 118.7$\pm7.8$ & 122.9$\pm8.2$\\
Companion offset ($\rm{mas}$) & - & - & 104$\pm{2}$ \\
Companion position angle $(\degree)$ & - & - & 288.6$\pm4.6$ \\
Companion contrast (mag) & - & - & 3.97$\pm{0.05}$ \\   
Flux Ratio (7.8)         & 7.8      &     7.8                     &  7.8       \\

\hline
Reduced \chisquare           & 65.1 & 60.0 & 48.0\\
Reduced \chisquare~2016 data       & 68.3  & 61.0  & 52.6   \\
\hline
\hline
\multicolumn{4}{c}{\hd} \\
Parameter & Symmetric & Asymmetric & Companion$^*$ \\
\hline 
Dust to gas ratio          &8.88${\pm1.34}\times 10^{-4}$ &7.16${\pm0.63}\times 10^{-4}$ &6.81${\pm0.64}\times 10^{-4}$\\
$\delta_{d}$&6.14${\pm0.57}\times 10^{-4}$ &7.14${\pm0.47}\times 10^{-4}$ &7.17${\pm0.42}\times 10^{-4}$\\
$\delta_{1}$      &2.77${\pm0.38}\times 10^{-2}$ &2.02${\pm0.41}\times 10^{-2}$ &2.02${\pm0.38}\times 10^{-2}$\\
$r_{d}$ $(\au)$  &0.11${\pm0.01}$ &0.11${\pm0.01}$ &0.11${\pm0.01}$\\
$r_{d}$ $(\rm{mas})$& $0.96{\pm0.08}$ & $0.96{\pm0.08}$ & $0.96{\pm0.08}$\\
$r_{1}$ $(\au)$       &8.4${\pm0.4}$               &7.3${\pm0.5}$                &7.3${\pm0.4}$\\
$r_{1}$ $(\rm{mas})$& $74{\pm4}$ & $64{\pm4}$ & $64{\pm4}$ \\
$r_{2}$ $(\au)$        &15.4${\pm1.3}$                &13.5${\pm0.6}$                 &13.4${\pm0.6}$\\
$r_{2}$ $(\rm{mas})$& $135{\pm11}$ & $118{\pm5}$ & $118{\pm5}$ \\
Inclination $(\degree)$    &30.3${\pm1.0}$               &29.8${\pm0.4}$                 &29.5${\pm0.6}$\\
Position angle $(\degree)$ &12.6${\pm2.1}$              &15.0${\pm0.9}$                 &15.7${\pm1.5}$\\
Star offset ($\rm{mas}$) & - & 0.09$\pm{0.03}$ & 0.11$\pm{0.03}$\\
Star offset position angle $(\degree)$ & - & 15.0$\pm9.1$ & 7.4$\pm8.9$\\
Companion offset ($\rm{mas}$) & - & - & 131 (fixed) \\
Companion position angle $(\degree)$ & - & - & 2.6 (fixed)\\
Companion contrast (mag) & - & - & 5.7$\pm{0.7}$ \\
Flux Ratio (5.2)         &  5.2            & 5.2                & 5.2              \\
\hline
Reduced \chisquare                 & 5.91 & 5.58  & 5.51 \\
Reduced \chisquare~2016 data       &    44.1 & 42.8 & 45.7 \\
\hline
\end{tabular}}
\begin{minipage}{\linewidth}
\vspace{0.1cm}
\footnotesize{$^*$This model is not statistically significant. The additional parameters in this model are not justified by the data, but is included for completeness.}
\end{minipage}
\end{center}
\end{table*}

Naively, \emcee~temperature would be increased by the value of reduced \chisquare, which is equivalent to scaling error bars on the data pixels by $\sqrt{\chi^2_{\rm shot}}$ to give $\chi^2=1$ due to uncertainties in addition to the modelled shot noise. 
However, this does not take into account the highly correlated speckle noise, and would underestimate our final uncertainties. 
The speckle noise has a characteristic solid angle of correlated noise which is $\sim(\lambda/D)^2$, or $\sim N_{\rm ind}=70$ pixels.
Therefore, to account for the speckle noise we set the temperature in \emcee~($T_{\mathrm{MC}}$) as $\sim N_{\rm ind}\times\chi^2_{\rm shot}$ ($\sim10000$), where $\chi^2_{\rm shot}$ is the reduced chi-squared of the best fitting model.
Thus the $\chi^2_{\rm shot}$ of $\gg1$ we have in our results means that photon statistics of the target and background do not dominate the uncertainty.
We have not studied sufficient calibrator-calibrator pairs to re-define our uncertainties reliably (based on the noise from the speckles). 

The main purpose of this work was to fit the inner regions of the disc, rather than the whole disc. The interested reader is directed to Appendix \ref{sec:seds_dust} for an investigation of the SEDs of the discs.

\subsection{\irs} \label{res_irs}

\begin{figure*}
\centering
\includegraphics[width=1.0\textwidth]{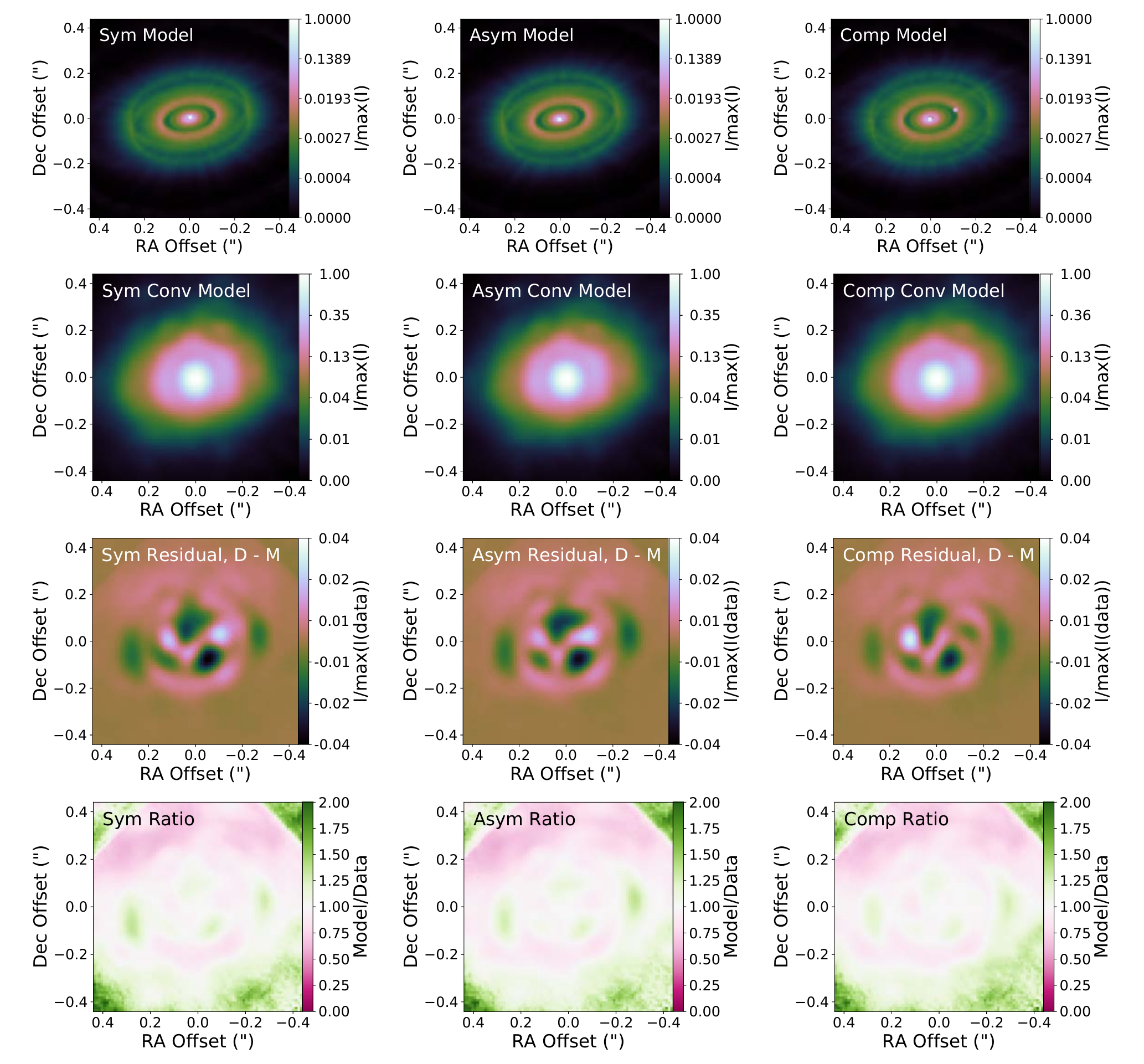}
  \caption{Converged results from the modelling of \irs. The left column shows the symmetric model results, the middle shows the asymmetric model results, and the right shows the companion model results. The top row is the model image for each model type. The second row is the convolved model. The middle row is the residual of the observed target data image minus the convolved model image. The fourth row is the ratio of the convolved model divided by the observed target data image. The bottom row is a plot of the midplane density profile of the disc.} 
  \label{irs48_models}
\end{figure*}

For \irs~the aim was to study the disc structure within $\sim20\,\au$ and also to see if a companion could be detected.
In our analysis of \irs, the 2015 data were used, with the best model for each of the model types for the 2015 data tested on the 2016 data.
Figure \ref{irs48_models} shows the resulting models, convolved images, residuals and ratio images.

One commonality shared by the \irs~models in their residual and ratio images in Figure \ref{irs48_models} is the two features (one on the left and one on the right of the disc) that are consistently present.
These `ear' features are green in the residuals and light green in the ratio images.
They are present in the same location as where the model appears brighter at around $30\,\au$ ($\sim\pm0.25\,$\arcsec), because of the overlap of the bright $\sim30\,\au$ ring.

This region of the disc is expected to become optically thick with opacity dominated by large grains, but our model only includes very small grains as we focus on the inner regions. 
This issue with the `ears' in the residual and ratio images could possibly be rectified by using a mixture of dust grains with varying sizes and compositions. 
There are some other features in the residual images that are consistent across all of the models, which may be due to an inadequate PSF library.
As mentioned above when uncertainties were discussed, this noise is also the reason for the high reduced \chisquare~values.

All three of the \irs~models have similar values for the parameters.
The symmetric and asymmetric models are most similar with nearly all parameters matching within uncertainties. 
The companion model is slightly different; as it must allow for the presence of the companion changing the brightness of the disc, thus adjusting the dust parameters and stellar position on accordingly.
The similarities and differences between the models are discussed in the following sections.

\subsubsection{Symmetric Disc}

The first column of Figure \ref{irs48_models} shows the results for the symmetric model.
The residual and ratio images show how well the model fits the data.
The L' band flux ratio of 7.8 is a match for the value we were fitting to.
The symmetric model of \irs~has a bright inner disc, $r_{1}$ at  $\sim112\,$mas ($\sim15\,\au$) and $r_{2}$ at $\sim238\,$mas ($\sim32.1\,\au$).
The location of $r_{1}$ is consistent with the location of the PAH emission modelled by \citet[][$\sim16\,\au$]{geers_spatially_2007}, and the rings suggested by \citet[][$\sim15\,\au$ and $\sim34\,\au$]{brown_matryoshka_2012}.

\subsubsection{Asymmetric Disc}

As mentioned before, the asymmetry in the disc was introduced by moving the star.
This introduces two new parameters that the symmetric model did not have, the star offset and position angle (generated from the $x$ and $y$ position of the star).
Because there are additional parameters a better fit is expected, and in this case a slightly better fit is achieved for the 2015 data.

Aside from the two new parameters, most of the parameter values are the same within uncertainties as for the symmetric model case.

\subsubsection{Asymmetric Disc with Companion}
\label{sec:irs_comp}

As mentioned earlier, a companion is added to the asymmetric model by including a second \radmc~point source.
The results for the planet model differ slightly from the asymmetric and symmetric model for the parameters relating to the dust structure. 
This is because having the second source causes some changes to the brightness of the disc.
The values of $r_{1}$ and $r_{2}$ are similar to those of the symmetric and asymmetric models, but $r_{d}$ has moved slightly further out.

The companion has a brightness contrast of 3.97$\pm{0.05}\,$mag with respect to the rest of the image. Note that this is a factor of $\sim$30 fainter than the disc emission or a factor of $\sim$5 fainter than the local disc emission within a diffraction limit ($\sim$80\,mas).
The position of our point-like asymmetry at $\sim104\,$mas and $\sim288\,\degree$ is consistent with one of the point sources found by \citet{Schworer17} at $\sim105\,$mas to the west of the star.
However, the \citet{Schworer17} object has a contrast of $\sim3.3\,$mag.

The reduced \chisquare~for this more complex model is the lowest of the three models for both 2015 and 2016, as reported in Table~\ref{tab:results_mean}.

\subsubsection{Summary of \irs~Results} 

The disc parameters are similar across all models for \irs, however there is improvement with the addition of an asymmetry and a companion.
Given this improvement, the small uncertainty in the companion radius and the narrow brightness contrast range, it is likely that there is an asymmetry in this region of the \irs~disc. 
However, we are unable to identify whether this is due to a disc asymmetry or a point-like companion at this stage.

In all cases the inner disc radius ($r_d$) is greater than what is considered to be the canonical inner disc for \irs.
\citet{Schworer17} find using imaging and SED fitting that there is likely no emission from the disc between 0.4 and $1\,\au$, and that the very small particles in the disc are present from $11\,\au$. 
We find that for the disc to be bright enough to match our observations, we do still require the inner disc, but present beyond $1\,\au$, with a second ring of emission present at $\sim15\,\au$ ($\sim112\,$mas).

Generally for \irs~the \chisquare~values are quite high.
Some part of this high \chisquare~is likely not only due to an inadequate PSF library, but  also due to insufficient complexity in the model, as evidenced by the negative residuals to the east and west at $\sim0.25\,\arcsec$ separation. 

\subsection{\hd} \label{res_hd}

\begin{figure*}
\centering
\includegraphics[width=1.0\textwidth]{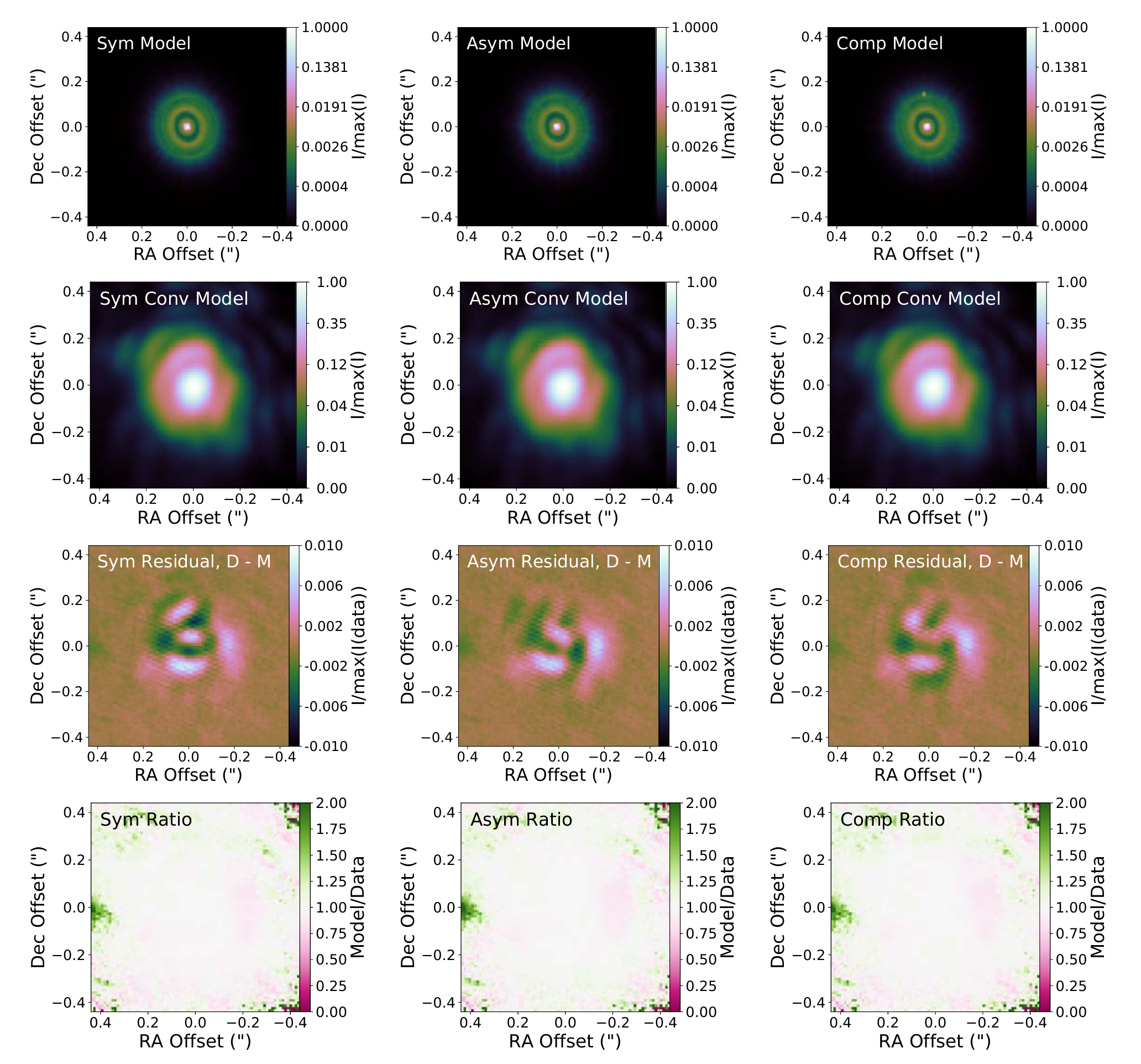}
  \caption{Results from the modelling of \hd. The layout is the same as for Figure \ref{irs48_models}.}
  \label{hd_models}
\end{figure*}

One goal of the study on \hd~was to see if we could detect the candidate companion proposed to be in the inner disc.
For \hd, the 2014 data were used for the analysis, with the best model for each of the model types for the 2014 data tested on the 2016 data.
The results of each model type are discussed below, and the figures corresponding to the model are shown in Figure \ref{hd_models}.
The same model types and figures are included here as were in the previous section (\ref{res_irs}) for \irs.

The inclination and position angle in all cases for the \hd~disc are different to those in the literature.
The inclination we find here is $\sim30\,\degree$, rather than the $\sim13\,\degree$ reported in \citet{panic_gas_2008}, and our position angle is $\sim15\,\degree$ compared to their $\sim30\,\degree$.
This difference may arise from the large difference in the spatial scales probed: the \citet{panic_gas_2008} observations using rotational lines of CO probed $>1\,$\arcsec separations compared to our $\sim0.1\,$\arcsec separations.
A disc warp, for example, could produce these slightly differing inner and outer disc geometries.

\subsubsection{Symmetric Disc}

For the symmetric model, the best fit has $r_{1}$ at $74\,$mas ($\sim8.4\,\au$), with $r_{2}$ at $135\,$mas ($15.4\,\au$). 
Previous studies of the disc found a ring at $\sim175\,$mas ($\sim20\,\au$), using modelling of the SED based on mid-IR observations \citep{honda_mid-infrared_2012} and imaging of large grains from 7mm observations \citep{osorio_imaging_2014}. 
Our deconvolved image of \hd~suggested a single ring at $\sim88\,$mas ($\sim10\,\au$), similar to that marginally detected by \citet[][$\sim100\,$mas]{ligi_investigation_2018}.
We attempted modelling \hd~with one wall, but found that the \chisquare~values were significantly lower when we used two walls.

The reduced \chisquare~for the 2014 data was 5.91 and for the 2016 data it was 44.1. 
There is more of a discrepancy between the results than there was for \irs, likely due to the very significant difference in the observing conditions in this case (flux varying by a factor of $\sim$2 during observations even after removing the most cloud-affected data) and this $\chi^2$ metric only taking shot noise into account (Section~\ref{sec:conv}).

\subsubsection{Asymmetric Disc}

The results for the asymmetric model are similar to the symmetric model.
The results fit the symmetric ones within a few standard deviations.
For this model $r_{1}$ is at $64\,$mas ($\sim7.3\,\au$), with $r_{2}$ at $118\,$mas ($\sim13.5\,\au$).
The location of the star to generate the asymmetry is $3-\sigma$ from the origin, suggesting that the star may not be off centre in \hd.
The reduced \chisquare~value for the asymmetric model is 5.58, which is a slight improvement over the symmetric model. 
The reduced \chisquare~of the 2016 data is 42.8 using the asymmetric model.

\subsubsection{Asymmetric Disc with Companion}
\label{sec:hd_comp}

The companion model has a similar density structure to both the symmetric and asymmetric models.
There is a wall at $64\,$mas ($7.3\,\au$), similar to the wall found in the symmetric and asymmetric models, as well as a wall at 118$\,$mas ($\sim13.4\,\au$).
When the companion position was freely explored, it was not detected in any location with certainty.
However, when the companion is placed at a location consistent with those found in the \citet{biller_enigmatic_2014} and \citet{reggiani_discovery_2014} studies (these are consistent with each other within the uncertainties), the \chisquare~was improved and the brightness contrast was found to be $5.7\pm{0.7}\,$mag. Our companion is at a lower significance than those papers, and not as significant as a general disc asymmetry.
Our brightness contrast is roughly consistent with the $6.5\pm0.5\,$mag found by \citet{reggiani_discovery_2014}, and the $\sim6.4\,$mag reported by \citet{biller_enigmatic_2014}. This point-like asymmetry is a factor of $\sim$100 fainter than the disc emission or a factor of $\sim$20 fainter than the local disc emission within a diffraction limit ($\sim$80\,mas).

For the 2016 data the reduced \chisquare~is 45.7, which is slightly higher than for the asymmetric model.

\subsubsection{Summary of \hd~Results}
In summary the results of the \hd~models are consistent with each other, and indicate the presence of an asymmetry in the disc.
We are able to recover the previously detected companion-like asymmetry, but are unable to determine the cause of the asymmetry at this stage. 
We also detect symmetric rings of dust at $\sim61\,$mas ($\sim7\,\au$) and $\sim114\,$mas ($\sim13\,\au$).
Our rings are at a similar location to the $\sim100\,$mas ($\sim11\,\au$) ring and asymmetry found by \citet{ligi_investigation_2018}.
The location of our inner disc ($r_d$) is at $0.11\,\au$ (0.96$\,$mas), which is slightly beyond the radius of $0.07\,\au$ found by \citet{chen_study_2018} from their modelling of the inner disc.

The substantial difference between the \chisquare~values for the 2014 and 2016 epochs for \hd~is due to the large variation in observing conditions between epochs. As discussed in Section \ref{sec:datred}, the observing conditions for the 2016 epoch were affected by weather, particularly \hd.

\subsection{Additional Companions}

\begin{figure}
    \centering
    \includegraphics[width=0.5\textwidth]{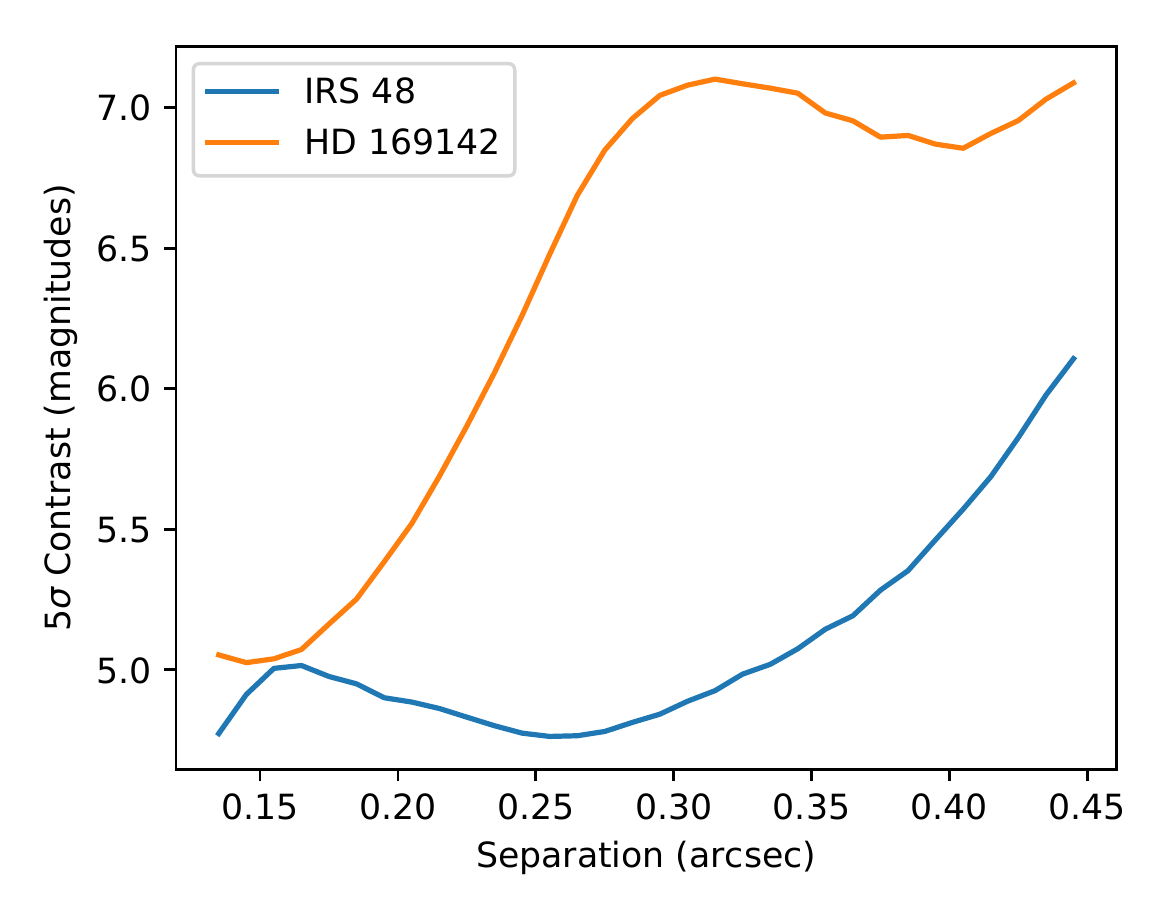}
    \caption{Detection limits as companion flux divided by total flux for additional companions fitted to the residuals of our disc models. These limits are not adequate to detect the majority of additional potential companions suggested in the literature.}
    \label{contrast_plot}
\end{figure}

We attempted to fit additional wide companions to our model residuals (i.e. companions present in addition to the companions fit in Sections \ref{sec:irs_comp}~and \ref{sec:hd_comp}), assessing the significance of companion fits by comparing fitted companion brightness to the root-mean square azimuthally averaged contrast as a function of separation. 
No additional companions were detected, and the resulting contrast limits are shown in Figure~\ref{contrast_plot}, with contrasts between 5 and 7 magnitudes with respect to the system absolute L' magnitudes of $\sim-0.6\,$mag for \irs~and $\sim0.6\,$mag for \hd.
The relatively poor contrast for \irs~is likely due to the inadequacy of our model in fitting the disc structure, resulting in relatively high residuals.
Additional companions would be detectable if they exist in the space below the lines in Figure \ref{contrast_plot}.

The possible 3.5\,$\rm{M_J}$ object at $40\,\au$ ($0.3\,\arcsec$) discussed in \citet{Schworer17}, deduced from disc structure and not from direct planetary emission, would then not be detectable in our data, unless it had a very high accretion rate of $\sim 3 \times 10^{-5}\, \rm{M_{J}}\, \rm{yr}^{-1}$ according to the models of \citet{Zhu15}. 
Note that our companion model is consistent with the asymmetry seen by \citet{Schworer17} which is interpreted as an asymmetric ring.
Potential companions aside from those near the location of the \citet{biller_enigmatic_2014}, \citet{reggiani_discovery_2014} and \citet{ligi_investigation_2018} candidates in \hd~include those suggested by \citet{perez_dust_2019}, which is at beyond $0.5\,$\arcsec; \citet{gratton_blobs_2019}, at a distance of $0.335\,$\arcsec, but a contrast of $10.1\,$mag; and \citet{pohl_circumstellar_2017}, one of which is at our inner limit and the other beyond our limit; thus none of these would be detectable with this method.

\section{Conclusions} \label{sec:conclusion}

Here we introduced our method for the study of transitional discs.
The method utilises images of target and calibrator stars taken with a thermal IR filter behind adaptive optics to reveal detail about the structures of their discs.
We use our new MCinMC (see Section \ref{sec:models}) code to make models of the observed discs, then convolve the models with calibrator images, and finally compare the convolved models with the data to determine how well the model fits our data.
The key conclusions we draw from this work are summarised here:

\begin{enumerate}

\item{In contrast to the historical view of transitional discs, we find that the ``gap'' region in two Herbig Ae stars is radiatively dominated by  emission from very small ($\la5\,$nm) grains and PAHs. This has been suggested before for both objects, and this paper confirms their presence with spatially resolved detections.}

\item{We confirm previously reported brightness asymmetries \citep{biller_enigmatic_2014, reggiani_discovery_2014, Schworer17} in both \hd~and \irs~using our complementary technique. However, the detection of overall disc asymmetry (modelled as an offset central star) was more significant than a point source companion model. For both objects, the co-located disc emission (within a diffraction limit) was brighter than previously reported ``companions'' by a factor of $\sim$5 or more, leading us to conclude that there is no need to invoke companions as anything more than a modelling convenience to explain the asymmetric emission.}

\end{enumerate}

The structure we detect in these discs suggests that the most common explanation of transitional discs having cleared inner holes may be too simple.
To better understand these transitional discs and the mechanisms within them, we need more complete models of the disc geometry and composition and an improved understanding of the PSF uncertainties in the observational data.

An improvement that could be made to our method would be to interpolate between the PSFs and to use a larger library, possibly incorporating multiple nights of observations or models of optical aberrations, in a similar way to LOCI \citep{lafreniere_new_2007} or KL eigenimages \citep{Soummer12}. 
These methods will ultimately be limited by the angular resolution of a single telescope, and improved spatial resolution with MATISSE \citep{Lopez14} or future concepts such as the Planet Formation Imager \citep[PFI,][]{ireland_status_2016} may be required to definitively distinguish between disc features and signs of exoplanets.

\section*{Acknowledgements}
We would like to thank the anonymous reviewer for their comments, which greatly improved the quality of this paper.

The data presented herein were obtained at the W. M. Keck Observatory, which is operated as a scientific partnership among the California Institute of Technology, the University of California and the National Aeronautics and Space Administration. The Observatory was made possible by the generous financial support of the W. M. Keck Foundation. The authors wish to recognize and acknowledge the very significant cultural role and reverence that the summit of Maunakea has always had within the indigenous Hawaiian community.  We are most fortunate to have the opportunity to conduct observations from this mountain.

E.K.B. would like to thank the Australian Government for their support through the Australian Government Research Training Program Stipend Scholarship and the Research School of Astronomy and Astrophysics at the Australian National University for the Masters of Astronomy and Astrophysics (Advanced) Scholarship.

M.J.I.~gratefully acknowledges funding provided by the Australian Research Council's Future Fellowship (FT130100235).

C.F.~gratefully acknowledges funding provided by the Australian Research Council's Discovery Projects (grants~DP150104329 and~DP170100603) and Future Fellowship Scheme (grant FT180100495), as well as the Australia-Germany Joint Research Cooperation Scheme (UA-DAAD).
The Monte Carlo simulations and data analyses presented in this work used high performance computing resources provided by the Australian National Computational Infrastructure (grant~ek9), and the Pawsey Supercomputing Centre with funding from the Australian Government and the Government of Western Australia, in the framework of the National Computational Merit Allocation Scheme and the ANU Allocation Scheme. We further thank for supercomputing resources at the Leibniz Rechenzentrum and the Gauss Centre for Supercomputing (grants~pr32lo, pr48pi and GCS Large-scale project~10391) and the Partnership for Advanced Computing in Europe (PRACE grant pr89mu).

S.K. acknowledges support from an ERC Starting Grant (Grant Agreement No. 639889) and STFC Rutherford Fellowship (ST/J004030/1).

These data were collected thanks to support from NASA KPDA grants (JPL-1452321, 1474717, 1485953, 1496788) and J.D.M. acknowledges NSF AST. 1311698.

This work has made use of data from the European Space Agency (ESA) mission {\it Gaia} (\url{https://www.cosmos.esa.int/gaia}), processed by the {\it Gaia} Data Processing and Analysis Consortium (DPAC, \url{https://www.cosmos.esa.int/web/gaia/dpac/consortium}). Funding for the DPAC has been provided by national institutions, in particular the institutions participating in the {\it Gaia} Multilateral Agreement.

\appendix

\label{sec:appendix}

\section{Parameter Convergence Study} \label{sec:conv_params}

\begin{figure}
        \centering
	\includegraphics[width=0.45\textwidth]{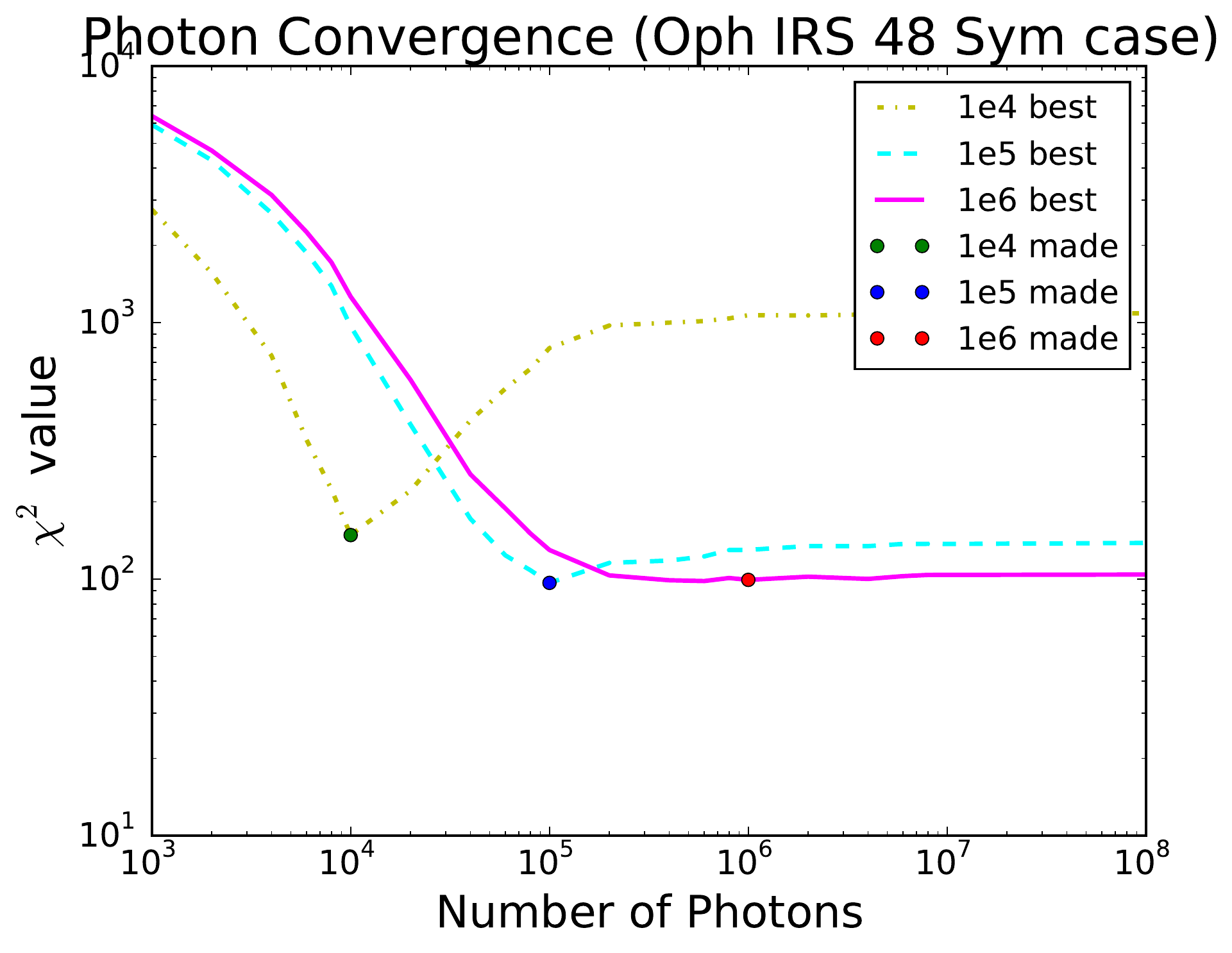}
    \caption{Convergence plots for different numbers of photons. The yellow dot-dashed line is the model that was found to be the best using $10^4$ photons, tested at different numbers of photons. The cyan dashed line is the model that was found using $10^5$ photons, tested at different numbers of photons. The solid magenta line was found using $10^6$ photons, and is the one chosen to be used in the modelling. The dots show where each different model for the numbers was found to have the best \chisquare.}
    \label{phot_conv}
\end{figure}

A convergence test was completed to determine the optimal number of photon packets to use in the model, so that it was stable, but not too computationally expensive. 
It was found that $10^6$ photon packets was good choice for the \radmc~thermal Monte Carlo to converge.
As seen in the plot in Figure \ref{phot_conv} there is a minimum for reduced \chisquare~for the best model found with $10^4$ or $10^5$ photon packets, because increasing the number of photon packets changed the best fit parameters.

\begin{table*}
\caption{Values used for the spatial grid in \radmc.}
\begin{center}
{\footnotesize\begin{tabular}{ccc}\hline
Variable & Grid Limits & Number of points between Grid Limits \\
\hline 
$\mathrm{r}$ ($\au$) & $\mathrm{r_{dust}}$, $\mathrm{r_{1}}$, $\mathrm{r_{1}+0.1}$, $\mathrm{r_{2}}$, $\mathrm{r_{2}\times1.1}$, 100   & 5, 20, 30, 20, 40 \\
$\mathrm\theta$ & 0, $\pi/3.$, $\pi/2.$, $2\pi/3.$, $\pi$ & 10, 30, 30, 10\\
$\mathrm\phi$ & 0, 2$\mathrm\pi$ & 60 \\
\hline

\end{tabular}}
\end{center}
\label{tab:grid}
\end{table*}

The convergence test used half and double the number of points in each parameter for the number of grid points. 
The result was that these different grids give a very similar \chisquare~values to the one shown in Table \ref{tab:grid}.

\section{Parameters for the companion brightness} \label{app:rad}
In all cases the companion had a temperature of 3500K and a mass of $10^{-3}\,\msolar$, these were chosen due to modelling constraints.
The brightness of the companion was adjusted by changing its radius, with a larger radius making the companion brighter. 
There was a lower limit on the radius of the companion, $0.02\,\rsol$, which had a negligible amount of flux, and we found that this was not important to the determination of the log likelihood.

\section{Additional \radmc~Parameters} \label{app:extra}

\begin{table*}
\caption{Set parameters for all models, most are defaults that \radmc~chooses, from the `problem\_params.inp' file.}
\label{table:extra_params}
\begin{center}
\footnotesize{\begin{tabular}{cc}\hline
Parameter & Value \\
\hline
Continuous Stellar Source & False \\
Discrete Stellar Source & True \\
Co\-ordinate System & Spherical \\
Number of Points for Wavelength grid & 19, 50, 30 \\
Number of Points for Wavelength grid (SED) & 100, 100, 30\\
Bounds of wavelength grid (\micron) & 0.1, 7.0, 25.0, 10000 \\
Bounds of wavelength grid (\micron) (SED) & 0.1, 1.5, 25.0, 10000 \\
Number of refinement levels & 3 \\
Number of the original grid cells to refine & 3 \\
Number of grid cells to create in a refinement level & 3 \\ 
Bulk Density of materials ($\mathrm{g}\mathrm{cm^{-3}}$) & 3.6, 1.8 \\
Grain size distribution power exponent & -3.5 \\
Maximum grain size & 10.0 \\
Minimum grain size & 0.1 \\
Mass fractions of the dust components to be mixed & 0.75, 0.25 \\ 
Number of grain sizes & 1 \\
Use finite size of star & no - take as point source \\
Modified Random Walk & Off \\
Number of photons for image generation & $2\times10^4$ \\
Number of photons for SED generation & $1\times10^5$ \\
Output format for \radmc~files & ASCII \\
Scattering mode & isotropic \\
Dust Temperature equal Gas Temperature & yes \\
Background Density ($\mathrm{g}\mathrm{cm^{-3}}$) & $1\times10^{-30}$ \\
Pressure scale height at innermost radius ($\au$) & 0.0\\
Reference radius at which $\mathrm{H_p}$/R is taken ($\au$) & 100. \\
Ratio of the pressure scale height over radius at reference radius for $\mathrm{H_p}$/R & 0.1 \\
Flaring index & 1./7. \\
Power exponent of the surface density distribution as a function of radius & -1.0 \\
Outer boundary of the puffed-up inner rim in terms of innermost radius & 0.0\\
Outer radius of the disc $\au$ & 100.\\
Surface density at outer radius of the disc & 0.0 \\
Surface density type & polynomial \\
Outer boundary of the smoothing in the inner rim in terms of innermost radius & 1.0\\
Power exponent of the density reduction inside of the the inner rim smoothing & 0.0\\
\hline

\end{tabular}}
\end{center}
\end{table*}

Geometric parameters for the spatial grid are shown in Table~\ref{tab:grid}, and additional parameters that were used (many of which did not deviate from the default \radmc~values) are shown in Table \ref{table:extra_params}.

\section{Spectral Energy Distributions and Dust Types}
\label{sec:seds_dust}

\begin{figure*}
\centering
  \resizebox{\textwidth}{!} {\begin{tabular}{@{}p{0.5\textwidth}p{0.5\textwidth}@{}}
    \includegraphics[width=.5\textwidth]{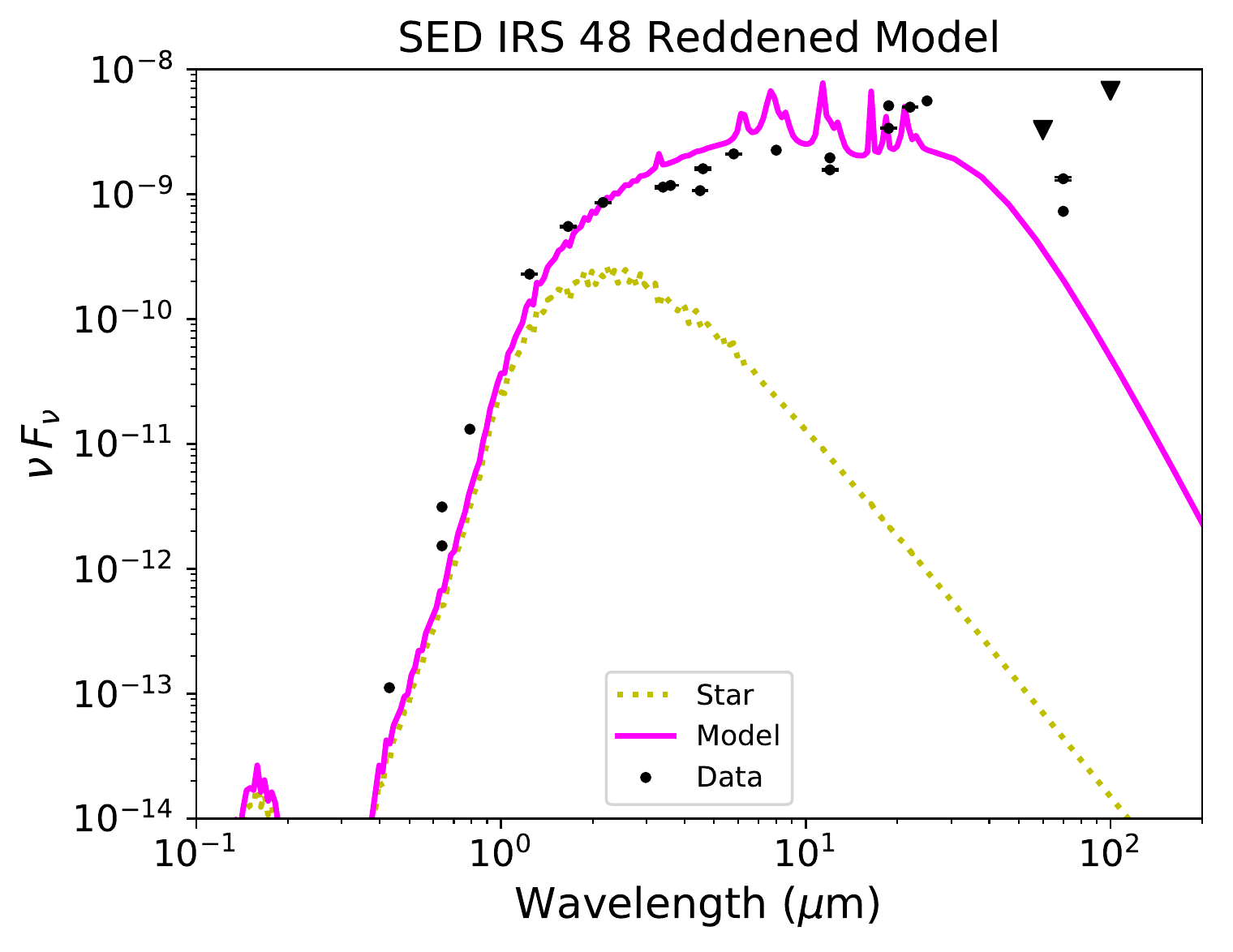} &
    \includegraphics[width=.5\textwidth]{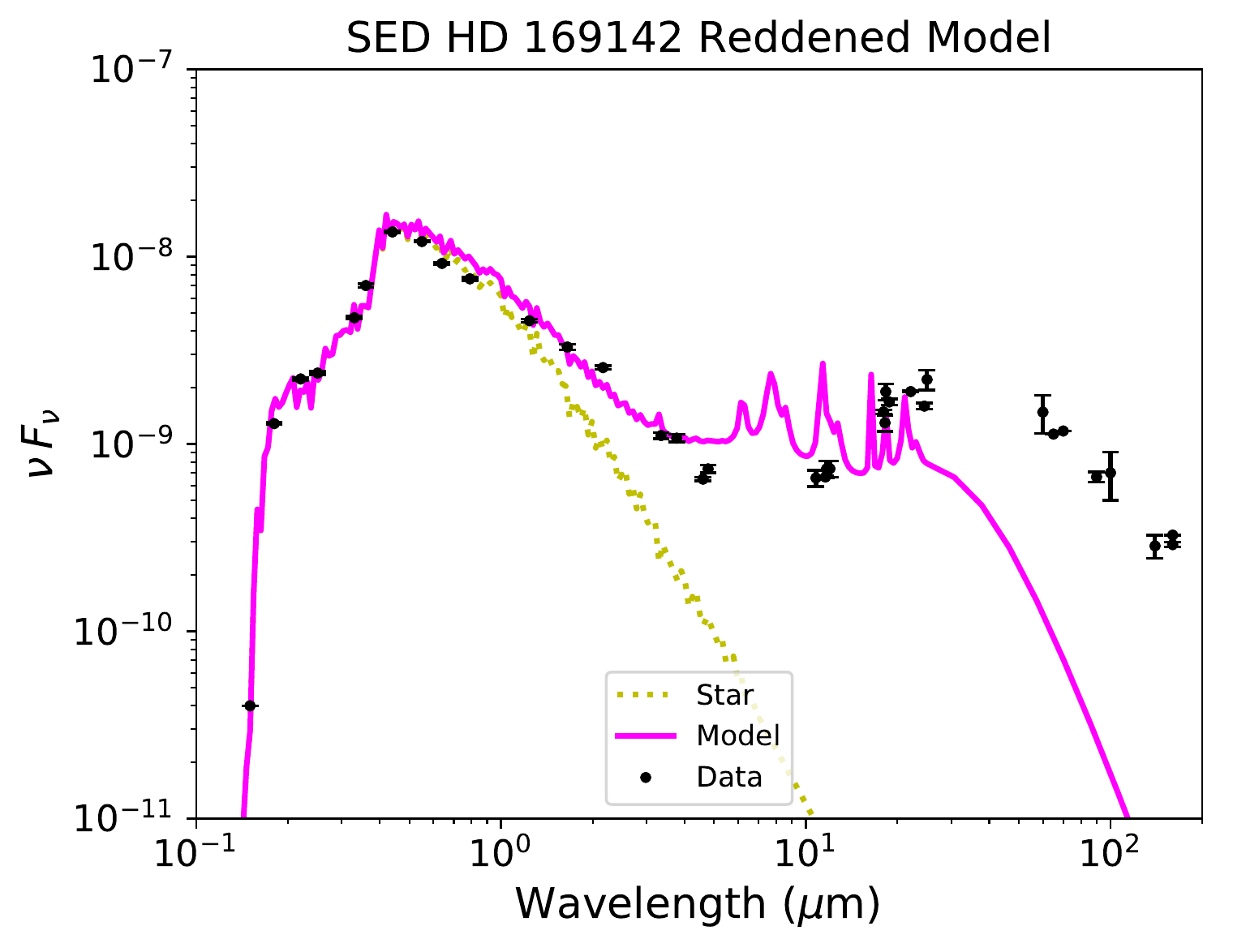} \\
    \end{tabular}}
  \caption{Shown here are the SEDs for the symmetric models in Table \ref{tab:results_mean}. Left is \irs~with the photometry data and a reddened model, and right is \hd~with the photometry data and a reddened model. The model does not fit all of the data points at all wavelengths (especially not in the un-modelled outer disc), but it does replicate the general shape of the observed SED. The black dots are the photometric data taken from literature, and the black triangles represent upper limits, see Table \ref{sed_table} for the references associated with these data.} 
  \label{both_sed}
\end{figure*}

The SED of the symmetric disc model for \irs~(left) and \hd~(right) are shown in Figure \ref{both_sed}.
The SED models are reddened to match the photometry of each of the objects, and are included to show that without explicitly trying to fit the SED we are able to replicate the shape.
The references for the data used in the SEDs can be found in Table \ref{sed_table}.

For \irs~it is difficult to de-redden the photometry or redden the model, because there is so much extinction towards it.
We use the \citet{cardelli_relationship_1989} reddening laws with $R_V=6.5$ and $A_V=12.9$, as was found in \citet{Schworer17}.
To calculate the level of reddening that was to be applied we used the extinction package for python by \citet{barbary_extinction_2016}.

We note that the SED is not fit very well by our model, and this is due to the constraints we have placed on ourselves for modelling the disc, as well as the large amount of extinction towards \irs.
A better fit could be achieved by differing dust types with radius, or with a different model configuration.
As fitting the SED was not the key goal of this work, but fitting the Keck L' filter data was, we leave detailed fitting of the SED for future work.

The SED of the symmetric model for \hd~(right of Figure \ref{both_sed}) is able to recover the flux at our target wavelength, but does not do well beyond this, particularly not in the un-modelled cool outer disc region.

The model was reddened using the same methods as for \irs, but with 
$R_V=3.1$ and $A_V=0.31$.
A better fit could be achieved with a different model configuration, specifically one that allows for varying the dust chemistry and distribution of the disc.

\begin{table}
\caption{References for SED data for \irs~and \hd.}
\begin{center}
\footnotesize{{\begin{tabular}{cc}\hline
\multicolumn{2}{c}{\irs} \\
Wavelength ($\micron$) & Reference \\
\hline
0.43, 0.64 & \citet{zacharias_vizier_2005} \\
0.64, 0.79 & \citet{erickson_initial_2011} \\
1.24, 1.66, 2.16  & \citet{cutri_vizier_2003} \\
3.4, 4.6, 12, 22 & \citet{wright_wide-field_2010} \\
3.6, 4.5, 5.8, 8, 70 & \citet{van_kempen_nature_2009} \\
12, 25, 60 ,100 & \citet{helou_infrared_1988} \\
18.7 & \citet{yamamura_vizier_2010}  \\
18.7 & \citet{geers_spatial_2007} \\
70 & \citet{fedele_digit_2013} \\
\hline
\hline
\multicolumn{2}{c}{\hd} \\
Wavelength (\micron) &  Reference \\
\hline
0.15, 0.18, 0.22, 0.25, 0.33 & IUE archival data\\
0.36, 0.44, 0.55, 0.64, & \citet{sylvester_optical_1996} \\
0.79, 3.77, 4.78 &  \\
1.24, 1.65, 2.16 & 2MASS All-sky Point \\
 &  Source Catalog \\
3.35, 4.6, 11.6, 22.1 & WISE All-sky Data Release  \\
 &  Catalog \\
10.8, 18,2 & \citet{jayawardhana_mid-infrared_2001} \\
11.7, 18.3  &  \citet{marinas_high-resolution_2011} \\
12, 25, 60, 100 & IRAS Point Source Catalog \\
18  & AKARI/IRC All-sky Survey Point \\
  & Source Catalog \\
  & \citep{ishihara_akari/irc_2010} \\
18.8, 24.5  & \citet{honda_mid-infrared_2012}  \\
65, 90, 140, 160 & AKARI/FIS All-sky Survey \\
 & Point Source Catalog \\
 & (Version 1.0)\\
70, 160  & \citet{meeus_gas_2010}  \\
\hline

\end{tabular}}}
\end{center}
\label{sed_table}
\end{table}

\begin{figure}
\centering
    \includegraphics[width=.5\textwidth]{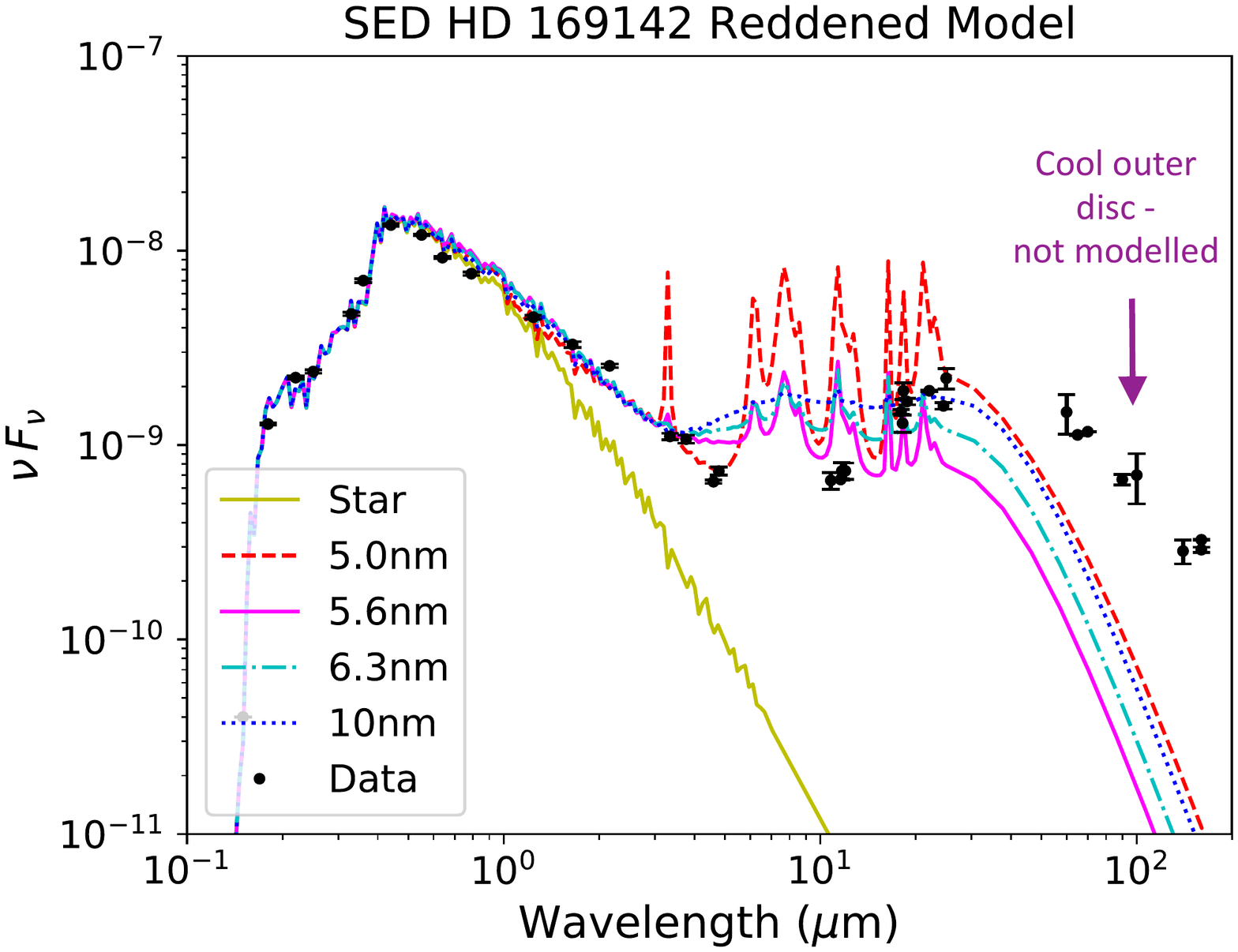} 
  \caption{Shown here are the SEDs for the symmetric disc model of \hd~with different dust types. The solid yellow line is the model of the stellar photosphere. The dashed red line is for dust that has a $5.0\,$nm grain-size, the solid pink line for dust that has $5.6\,$nm grains, the dot-dashed cyan line is for dust with a grain-size of  $6.3\,$nm and the dotted blue line is for dust that is $10.0\,$nm. The purple arrow indicates the cool outer disc, which we are not attempting to fit here.}
  \label{hd_dust}
\end{figure}

We also tested \hd~with a few different dust sizes to investigate how grain-size changed the fit to the SED.
The results of these tests are found in Figure \ref{hd_dust}.
Each dust type was started from the best fit with the $5.6\,$nm dust and then the MCinMC code was run to find a best fit model for the new dust type.
The grain-sizes investigated were $5.0\,$nm, $6.3\,$nm, and $10.0\,$nm, with the $5.6\,$nm dust to compare to.
The composition of the dust is all the same carbon and PAH dust mix as discussed above from the \citet{draine_infrared_2001, draine_infrared_2007} set of neutral carbon and PAH dust.

Dust smaller that the one used for our analysis has very strong emission from PAHs, whereas dust of a larger grain size has weaker PAH emission. 
Larger dust is too cool to recover the features we would like to replicate.
The chosen dust size of $5.6\,$nm allows us to have PAH features with some smoothing due to graphite.
Using a significantly smaller dust would move into the realm of needing quantum heating of the dust particles, which would be difficult to parameterise for our model, so we chose to use dust models readily available from the literature.

\bibliographystyle{mnras}
\bibliography{Birchall_paper1_arxiv}{}


\bsp	
\label{lastpage}
\end{document}